\documentclass[a4paper,11pt]{article}
\usepackage{aaskaiid}
\usepackage{orcidlink}

\usepackage{enumitem}
\setlist[itemize]{noitemsep, topsep=0pt}

\newcommand \hi{{H\textsc{i}}}
\newcommand{\MHz}{\ensuremath{\textrm{ MHz}}}
\newcommand{\GHz}{\ensuremath{\textrm{ GHz}}}
\newcommand{\EM}{\ensuremath{\textrm{EM}}}
\newcommand{\pccmsix}{\ensuremath{\textrm{ pc cm}^{-6}}}

\newcommand{\radmsq}{\ensuremath{\textrm{ rad m}^{-2}}}
\newcommand{\cucm}{\ensuremath{\textrm{ cm}^{-3}}}
\newcommand{\uG}{\ensuremath{\, \mu \textrm{G}}}

\title{The Multi-phase HI of the Milky Way and Nearby Galaxies}
\ShortTitle{The Multi-phase HI of the Milky Way and Nearby Galaxies}

\author[1]{Marc-Antoine Miville-Desch\^enes\orcidlink{0000-0002-7351-6062}}
\ShortName{M.-A. Miville-Desch\^enes et al.} 
\author[2,3]{J. R. Dawson\orcidlink{0000-0003-0235-3347}}
\author[4]{Narendra Nath Patra}
\author[1]{Erwan Allys\orcidlink{0000-0003-3755-7593}}
\author[5]{Prerana Biswas}
\author[1]{Frances Buckland-Willis\orcidlink{0000-0003-4213-8094}}
\author[6,7]{Susan E. Clark\orcidlink{0000-0002-7633-3376}}
\author[15]{James Dempsey\orcidlink{0000-0002-4899-4169}}
\author[8]{John Dickey\orcidlink{0000-0002-6300-7459}}
\author[9]{Adriana Gazol}
\author[1]{Benjamin Godard\orcidlink{0000-0003-0060-8887}}
\author[10]{Patrick Hennebelle\orcidlink{0000-0002-0472-7202}}
\author[11,12]{Alex S. Hill\orcidlink{0000-0001-7301-5666}}
\author[13,14]{Min-Young Lee\orcidlink{0000-0002-9888-0784}}
\author[6,7]{Minjie Lei\orcidlink{0000-0002-2679-4609}}
\author[15]{Callum Lynn\orcidlink{0000-0001-6846-5347}}
\author[1]{Antoine Marchal\orcidlink{0000-0002-5501-232X}}
\author[15]{Naomi McClure-Griffiths\orcidlink{0000-0003-2730-957X}}
\author[16]{Claire Murray\orcidlink{0000-0002-7743-8129}}
\author[15]{Van Hiep Nguyen\orcidlink{0000-0002-2712-4156}}
\author[6,7]{Marta Nowotka\orcidlink{0009-0002-0282-4188}}
\author[17]{Theo J. O'Neill\orcidlink{0000-0003-4852-6485}}
\author[16]{Josh E. G. Peek\orcidlink{0000-0003-4797-7030}}
\author[18,19]{Nickolas M. Pingel\orcidlink{0000-0001-9504-7386}}
\author[20]{Mary Putman\orcidlink{0000-0002-1129-1873}}
\author[18]{Daniel Rybarczyk\orcidlink{0000-0003-3351-6831}}
\author[21]{Nirupam Roy}
\author[15]{Amit Seta\orcidlink{0000-0001-9708-0286}}
\author[21]{Sarkar Sougata\orcidlink{0009-0004-4965-8548}}
\author[22]{Juan Diego Soler\orcidlink{0000-0002-0294-4465}}
\author[9]{Enrique Vazquez-Semadeni\orcidlink{0000-0002-1424-3543}}
\author[23]{Trey V. Wenger\orcidlink{0000-0003-0640-7787}}
\author[17]{Catherine Zucker\orcidlink{0000-0002-2250-730X}}

\affiliation[1]{Laboratoire de Physique de l’Ecole normale supérieure, ENS, Université PSL, CNRS, Sorbonne Université, Université Paris-Diderot, Sorbonne Paris Cité, Observatoire de Paris, Paris,
France}
\affiliation[2]{School of Mathematical and Physical Sciences and Astrophysics and Space Technologies Research Centre, Macquarie University, North Ryde 2109, Australia}
\affiliation[3]{Australia Telescope National Facility, CSIRO Space \& Astronomy, Epping 1710, Australia}
\affiliation[4]{Indian Institute of Technology Indore, India}
\affiliation[5]{Indian Institute of Astrophysics, Bengaluru, India}
\affiliation[6]{Department of Physics, Stanford University, Stanford, CA 94305, USA}
\affiliation[7]{Kavli Institute for Particle Astrophysics \& Cosmology, P.O. Box 2450, Stanford University, Stanford, CA 94305, USA}
\affiliation[8]{School of Natural Sciences, Private Bag 37, University of Tasmania, Hobart, TAS 7001, Australia}
\affiliation[9]{Instituto de Radioastronomía y Astrofísica, Universidad Nacional Autónoma de México, Apdo. Postal 3-72, Morelia, Michoacán, Mexico}
\affiliation[10]{Université Paris-Saclay, Université Paris Cité, CEA, CNRS, AIM, 91191 Gif-sur-Yvette, France}
\affiliation[11]{Department of Computer Science, Math, Physics, \& Statistics, University of British Columbia, Okanagan Campus, Kelowna, BC V1V 1V7, Canada}
\affiliation[12]{Dominion Radio Astrophysical Observatory, Herzberg Research Centre for Astronomy and Astrophysics, National Research Council Canada, PO Box 248, Penticton, BC V2A 6J9, Canada}
\affiliation[13]{Korea Astronomy and Space Science Institute, 776 Daedeok-daero, Daejeon 34055, Republic of Korea}
\affiliation[14]{Department of Astronomy and Space Science, University of Science and Technology, 217 Gajeong-ro, Daejeon 34113, Republic of Korea}
\affiliation[15]{Research School of Astronomy and Astrophysics, The Australian National University, Canberra, ACT 2611, Australia}
\affiliation[16]{Space Telescope Science Institute, 3700 San Martin Drive, Baltimore, MD 21218, USA}
\affiliation[17]{Center for Astrophysics | Harvard \& Smithsonian, 60 Garden Street, Cambridge, MA 02138, USA}
\affiliation[18]{University of Wisconsin–Madison, Department of Astronomy, 475 N Charter Street, Madison, WI 53703, USA}
\affiliation[19]{Department of Astronomy, Indiana University, 727 East Third Street, Bloomington, IN 47405, USA}
\affiliation[20]{Department of Astronomy, Columbia University, New York, NY 10027, USA}
\affiliation[21]{Indian Institute of Science, Bengaluru, India}
\affiliation[22]{Istituto di Astrofisica e Planetologia Spaziali, INAF, Via Fosso del Cavaliere 100, I-00133 Roma, Italy}
\affiliation[23]{California State University, Chico, 400 W. First Street, Chico, CA 95926, USA}

\abstract{
Atomic hydrogen (\hi) is the dominant baryonic component of the interstellar medium (ISM) in Milky Way–like galaxies and the reservoir from which molecular clouds and stars ultimately form. The condensation of diffuse \hi\ into cold structures is governed by a complex interplay between radiative cooling, turbulence, magnetic fields, stellar feedback, and galactic dynamics, acting over scales ranging from astronomical units to kiloparsecs. Understanding how these processes regulate the thermal structure of the \hi, the formation of cold clouds, and the transfer of matter and energy across scales is essential for connecting the small-scale physics of the ISM to the evolution of galaxies. Recent advances from SKA precursors have transformed our view of the atomic ISM, revealing a highly structured and filamentary cold medium, increasing the density of \hi\ absorption measurements by orders of magnitude, and enabling new approaches to infer the thermodynamic and magnetic properties of the gas from spectral-line datasets. SKA-mid will provide the first comprehensive characterization of \hi\ as a multi-phase, turbulent, and magnetized medium across the Milky Way and nearby galaxies. Its combination of sensitivity, angular resolution, spectral resolution, and survey speed will enable matched emission–absorption studies, dense optical-depth grids, and detailed mapping of the atomic-to-molecular transition over a broad range of environments. Combined with polarization, Zeeman, recombination-line, and multi-wavelength observations, SKA-mid will establish a unified observational framework to study the evolution of diffuse matter in galaxies, in connection with star formation, from the Solar neighborhood to galactic scales.
}

\begin{document}

\maketitle

\section{Introduction}

For the past 75 years the $21\,$cm emission line of \hi\ has been a key tracer of galaxy structure and evolution, revealing the presence of dark matter through galaxy rotation curves, the ubiquity of atomic hydrogen in galactic disks and halos, the impact of stellar feedback on interstellar matter through super-shells, and the multi-phase properties of the \hi\ with dedicated $21\,$cm absorption observations \citep{dickey1990,kalberla2009,McClure-Griffiths2023}. We have learned that \hi\ is composed of a mixture of small filamentary and cold clouds ($\sim 80\,$K) within a volume filling and magnetized warm medium ($\sim 6000\,$K). 
Most of the interstellar matter in Milky-Way type galaxies is in this multi-phase \hi, with atomic disks extending several times the molecular and stellar ones. The formation of \hi\ clouds, the seeds of the dense interstellar medium (ISM) where stars form, is the result of multi-scale non-linear physics including gas cooling and magnetised turbulence. 
This \hi\ condensation plays a dominant role in setting the star formation efficiency of galaxies, as much as the combination of gravity, turbulence and stellar feedback in molecular clouds \citep[e.g.][]{colman2025}. 

Until recently, this process, fundamental as it shapes the structure of matter, has been difficult to fully characterize due to observational limitations (sparse absorption measurements and the difficulty in separating cold and warm phases from $21\,$cm emission data). 
In addition, \hi\ evolution is not currently resolved in the cosmological simulations that aim to encompass the known laws of physics to describe the evolution of matter in the Universe over cosmic times. For computing reasons, these numerical experiments depart from fundamental laws at scales below $\sim 100\,$pc \citep[this is where the model breaks --][]{heiles2019} and rely on empirical relations deduced from observations (so-called sub-grid physics). Therefore, our understanding of the evolution of the Universe as a whole depends on the precision with which the physics of the cloud/core/star formation scenario and its related stellar feedback loop can be modelled on scales where it can be resolved, i.e. in nearby systems like the Milky Way and nearby galaxies. This fundamental and key aspect of our understanding of the evolution of matter in the Universe can not be overlooked. 

The evolution of diffuse interstellar matter in galaxies is complex, as it depends strongly on local physical conditions: the radiation field, magnetic field, metallicity, density, pressure, turbulent Mach number, and more. Exploring the \hi\ from the very local environment to nearby galaxies is a fundamental way to build a multi-scale / multi-process view of how energy and matter are transferred across scales through the interstellar turbulence cascade, how stellar feedback and gravitational disk instabilities shape matter, inject turbulent energy and contribute to the formation of clouds, and how magnetized interstellar turbulence is shaping this multi-phase medium at parsec scales and below. 

Over the past few years, several technical aspects have converged simultaneously, thanks in part to data from SKA-precursors like ASKAP, MeerKAT and LOFAR, providing a totally new perspective on the study of diffuse matter that will be magnified with SKA: (1) 21\,cm imaging at high resolution and high sensitivity in the Milky Way and nearby galaxies; (2) an increase by a factor of more than 100 in the 21\,cm absorption source density; (3) new advanced analysis methods to segment \hi\ phases from 21 cm data cubes; (4) reconstruction of the three-dimensional structure, kinematics and radiation field of the local ISM out to distances of up to 1--2 kpc; (5) effective coupling of numerical experiments to observations through realistic synthetic observations; (6) polarized synchrotron and Zeeman observations allowing new studies of the magnetic field in the \hi; (7) pulsar monitoring revealing very small scale \hi\ ($\sim$10--100 AU) and the dissipation of interstellar turbulence; (8) observations of so-called ``CO-dark'' molecular gas that reveal the \hi-H$_2$ transition. 

This new exploration space is opening a larger range of physical conditions and scales than has ever been probed before. Detailed imaging of the \hi\ physical properties in the Solar neighborhood\footnote{defined here as matter located within the local arm or the Milky Way, i.e., within 200-300\,pc of the Sun, also referred to as the Local Bubbble} will reveal the structure of cold clouds and how the magnetized turbulence cascade lead to their formation.
Moving away from the Solar neighborhood to nearby galaxies and their halos will provide a more global picture of these first steps of star formation. 
Such a broadening in scope enabled in the SKA era promises to shape a much more detailed astrophysical knowledge of baryonic matter evolution in galaxies, that will benefit our understanding of the cosmic evolution as a whole. 

This chapter of ``Advancing Astrophysics II: Preparing for Science with the SKAO'' builds on the foundations laid by recent reviews of Galactic \hi\ \citep{heiles2019,McClure-Griffiths2023,stanimirovic2024}, and also takes strong inspiration from \cite{McClureGriffiths2015aska} who imagined studies of ``Galactic and Magellanic Evolution with the SKA''. This chapter updates this vision in the light of recent results obtained with SKA precursors and announced performances of the upcoming SKA AA4. 
Sections 2 to 4 present the current state, challenges and open questions of the many aspects of \hi\ astrophysics. For all of them, Section~\ref{sec:SKA-observing-program} describes the leap forward that will be enabled by SKA. 

\section{Thermal state of the \hi: the cloud formation process in galaxies}

\begin{figure*}
    \centering
    \includegraphics[scale=0.42]{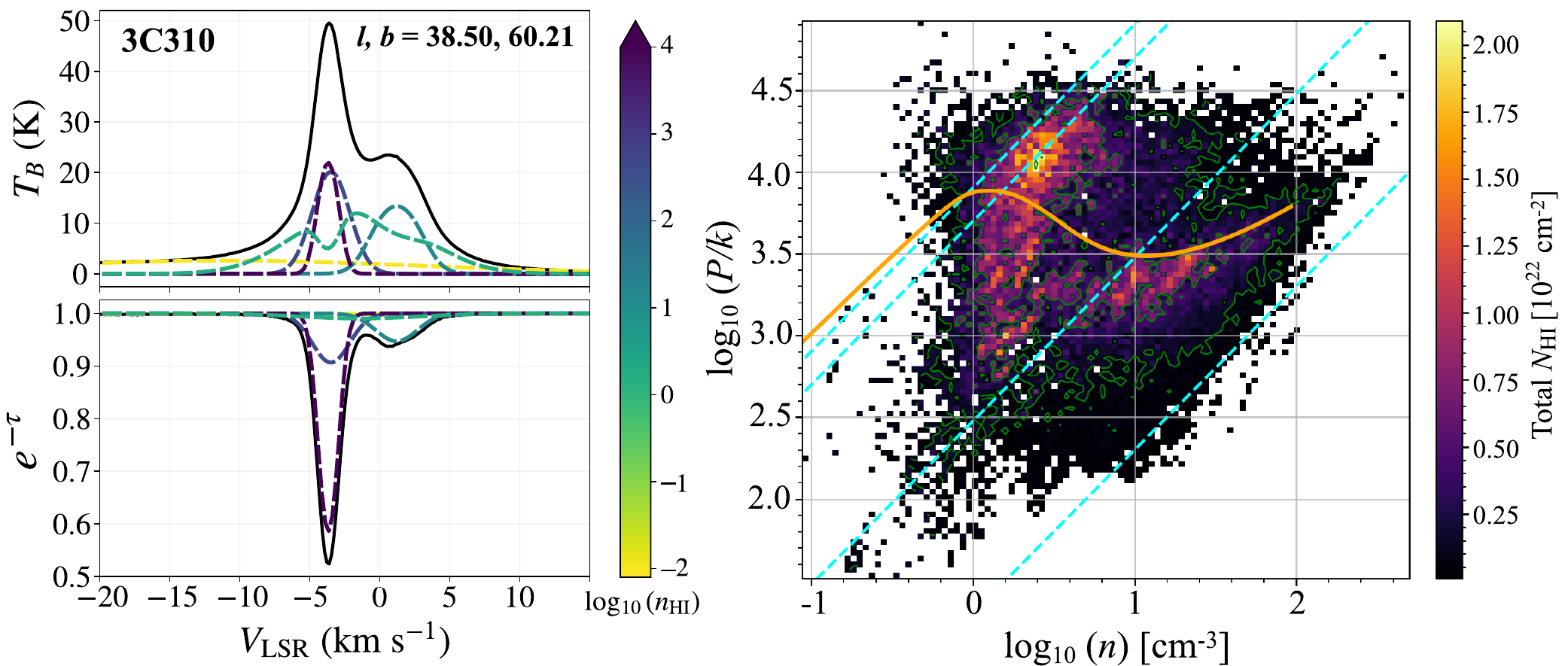}
    \caption{Example outputs from \texttt{caribou\_hi} \citep{wenger2024}. {\bf Left:} model fits to 21\,cm emission (top) and absorption (bottom) spectra \citep{McClure-Griffiths2023}. Colors indicate the fitted \hi{} volume density for each component. {\bf Right:} Pressure-density phase diagram constructed from the results of model fitting towards 462 GASKAP-\hi\ sightlines towards the LMC foreground (data originally presented in \citealt{nguyen2025}). Cyan lines denote lines of constant kinetic temperature bracketing the expected ranges for the WNM and CNM (5000 K to 8000 K, 20 K to 300 K). The orange line is the thermal equilibrium curve for the solar neighborhood adopted from \citet{wolfire2003}.}
    \label{fig:caribou_results}
\end{figure*}

Early after the detection of the \hi\ $21\,$cm line emission of the Milky Way in the 1950s, a clear difference appeared between lines-of-sight where the \hi\ is in emission (broad lines of several tens of km~s$^{-1}$ with narrower components) and others where absorption against bright background radio sources shows only the narrowest features of the Galactic \hi\ (see Fig.~\ref{fig:caribou_results}-left). This fundamental difference, observed everywhere on the sky, revealed the multi-phase nature of the neutral atomic gas. It was soon understood that, under the heating and cooling conditions of the ISM, a thermal instability is present that leads to a cloud condensation process \citep{field1965}. Any dynamical perturbation of the system, provided it is long enough compared to the cooling time, leads to a two-phase structure comprising the cold neutral medium (CNM) and warm neutral medium (WNM), with number densities, $n$, and kinetic temperatures, $T_k$, of ($n$, $T_k$) = (7--70 cm$^{-3}$, 60--260 K) and (0.2--0.9 cm$^{-3}$, 5000--8300 K), respectively \citep{mckee1977, wolfire2003}. Therefore, time-dependent and dynamical effects such as turbulence and supernova-driven shocks play a crucial role in shaping the thermal properties of \hi. These mechanisms are expected to modify the CNM formation process as well as maintaining a significant fraction of gas in the thermally unstable medium (UNM) with properties intermediate between the CNM and WNM \citep[e.g.,][]{audit2005,saury2014}.
These thermal properties of the \hi\ are part of the larger context of ISM phases; the \hi\ is a fundamental bridge between the ionized gas (Warm Ionized Medium, WIM, and Hot Ionized Medium, HIM) and the denser molecular gas, bright or dark in CO emission.

Decades of $21\,$cm observations of the Milky Way \hi, in absorption and emission, have confirmed this general picture; the cloud phase of the ISM starts out of this out-of-equilibrium condensation process, amplified later on by self-gravity for the densest structures. Understanding this fundamental process for star formation and the evolution of galaxies requires us to be able to reconstruct the thermal properties of the gas from $21\,$cm data. Two methods have been developed. The classical one uses absorption measurements on discrete lines-of-sight against background radio sources, combined with emission data from the immediate surroundings on-sky. More recently methods have been developed to extract thermal information from the blended, fully sampled, emission data alone. 

\subsection{Building the thermal picture with absorption studies}

\subsubsection{The Milky Way}
\label{sec:MW-absorption}

Determining the optical depth and spin temperature ($T_s$) of \hi\ --- essential for characterizing the gas thermodynamics --- is achieved by simultaneous measurements of both $21\,$cm emission and absorption.
As a result, most of our knowledge of the temperature of the neutral ISM of the Milky Way has been obtained from targeted observations of lines of sight against bright radio continuum sources, combined with matched emission data from very close positions on the sky.
As the opacity of the 21\,cm line depends on $1/T_s$, absorption measurements provide information on the cold gas (CNM) while the emission data contains both the CNM and WNM emission. The combination of absorption and emission around the source, followed by a careful multi-component Gaussian decomposition of the resulting spectra \citep{heiles2003}, provides local information on the thermal properties (Fig.~\ref{fig:caribou_results}).

We have learned that the CNM is ubiquitous, detected in $>80\%$ of Milky Way lines of sight, with $T_s\sim50$--$100\,$K \citep{heiles2003, murray2018b, stanimirovic2014, McClure-Griffiths2023}. CNM detected via \hi\ self-absorption is even colder, with $T_s\sim20$--$50\,$K \citep[see Section~\ref{sec:H2} and][]{denes2018}. The fraction of gas in the CNM versus the UNM and WNM (the {\em CNM fraction}) varies from very low (a few percent) in quiescent, local clouds \citep{murray2021} to around $20\%$ in the general local ISM \citep{McClure-Griffiths2023}, to even higher ($30$--$60$\%) in the environments of molecular clouds \citep{stanimirovic2014, nguyen2019}. These results also show that the CNM has a volume filling factor of $\sim 1\%$; it is a cloud phase immersed in the volume filling UNM and WNM.

Given their higher spin temperatures, detecting absorption signatures of WNM and UNM demands exceptional sensitivity to \hi\ optical depth. As a result, detections of \hi\ with $T_s>1000\rm\,K$ are relatively rare \citep{carilli1998, dwarakanath2002, murray2015}. From the highest-sensitivity searches for WNM absorption, a statistical detection through spectral stacking revealed that it has a characteristic $T_s\sim7200\rm\,K$ \citep{murray2014}. This is much warmer than predicted by numerical simulations, and implies that additional excitation via Ly$\alpha$ scattering is required to maintain such a high excitation temperature \citep{kim2014}. Direct detections of the UNM in absorption reveal that it has a significant mass fraction of $\sim20\%$ by mass \citep{murray2018b, nguyen2019}.

\subsubsection{The pathfinders revolution}

A key breakthrough enabled by SKA precursors and pathfinders such as ASKAP and MeerKAT is the capacity to perform untargeted wide-area surveys that simultaneously probe background continuum sources and spectral-line emission. This ability to carry out untargeted absorption surveys with sufficient sensitivity (now referred to as ``absorption grids'' due to their high density of background sources per square degree) is ushering in a new era for detailed studies of the temperature and optical depth structure of the neutral ISM.

With its first $\sim$100 hours of observing time, the GASKAP-\hi\ survey obtained 462 \hi\ absorption detections over 250 square degrees \citep{nguyen2025}, while \citet{lynn2025} revealed CNM and UNM-like components buried within the noise by stacking $\sim 2250$ GASKAP-\hi\ sources without individual \hi\ absorption detections at the $3\sigma$ level.
Similarly, the MeerKAT Absorption Line Survey \cite[MALS,][]{gupta2025} recently reported a total of 3640 \hi\ absorption detections over an area of $\sim$800 deg$^2$.
The sheer number of detections of these new surveys, several per square degrees, are far surpassing the total number of detections previously obtained by all past targeted surveys combined \citep[372 sources over the whole sky][]{McClure-Griffiths2023}. These early results from ASKAP and MeerKAT demonstrate the incredible potential of SKA-mid to produce dense absorption grids against which the thermodynamic properties of \hi\ can be revealed at unprecedented detail. This was foreseen in \citet{McClureGriffiths2015aska} as one of the backbones of an SKA-mid survey of Galactic \hi, and it remains even more true now with the clear demonstrations from the pathfinders. This is also true for \hi\ in external galaxies (see Section~\ref{sec:external-galaxies}).

\subsubsection{Radiative transfer modeling}

The increase in \hi\ absorption statistics has been accompanied by recent methodological advances. A noteworthy example, showcasing the potential of absorption grid to provide new insights on the thermal properties of \hi, is the Bayesian forward modeling approach developed by \citet{wenger2024}. 
Unlike traditional emission/absorption analysis, physical quantities and their uncertainties are not derived from Gaussian fits to the spectra, but are predicted by a physical model parameterized in terms of those properties. 
For each model cloud along a sight line (corresponding to a single Gaussian spectral component), this method produces predictions of the kinetic temperature, density, and pressure (both thermal and non-thermal) of the atomic gas, with robust constraints on their uncertainties, as well as the traditional measurements of $T_{\rm s}$, optical depth and column density. It can also account for different beam sizes in emission and absorption, and can deal self-consistently with self-absorption. 
Figure \ref{fig:caribou_results} shows an example of a phase diagram, $P/k$ versus $n$, constructed from the 462 GASKAP-\hi\ sight-lines discussed earlier. 
While the uncertainties on individual data-points can be very large, the ability to generate posterior predictive $P-n$ phase diagrams can directly constrain the ensemble thermal properties of the gas, providing direct tests of theoretical predictions. A future SKA-mid project would take full advantage of such methodological improvements.


\begin{figure*}
  \centering
  \includegraphics[width=0.9\linewidth]{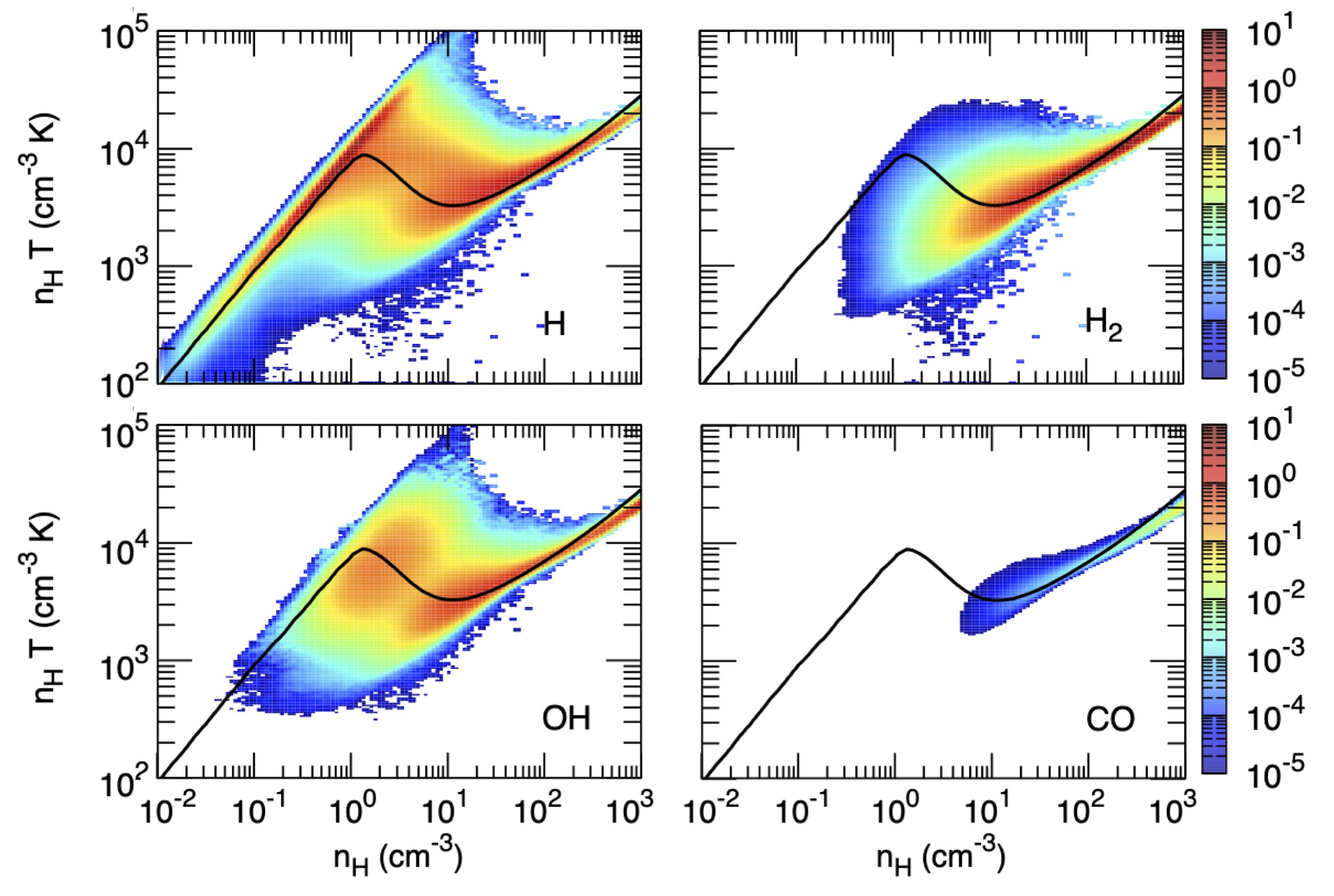}
  \caption{Predicted $P-n$ diagrams for \hi\, H$_2$, OH and CO, corresponding to the \hi-H$_2$ transition from diffuse to dense gas. These were constructed using models from \cite{bellomi2020,godard2023}. The \hi\ shows a typical two thermally stable branches (CNM and WNM) with significant gas in the thermally unstable regime. The OH and H$_2$ highlights the expected presence of molecules in the diffuse gas, with CO being concentrated in cold and dense structures. The color scale represents the normalized two-dimensional probability distribution functions of each species per $\log(n)$ and $\log(P)$. }
  \label{fig:P-n_HI-H2}
\end{figure*}

\subsection{The atomic to molecular transition}

\label{sec:H2}

With the ability to probe the CNM, we can study the transition from cool, atomic clouds to diffuse molecular gas (sometimes called ``CO-dark'' for its lack of detectable CO emission; e.g. \citealt{grenier2005}) and on to dense molecular clouds. On an integrated sightline basis through the local ISM, this transition occurs when the total H column density is between 10$^{20}$ and 10$^{22}$ cm$^{-2}$ \citep{bellomi2020}, with the concentration of H$_{2}$ thought to increase rapidly in small regions first, as the ambient UV is screened out by increasing dust column density \citep{park2023,liszt2025}. In nearby molecular clouds the SKA promises the potential to reveal the spatial and kinematic structure of the atomic-molecular transition on sub-parsec scales.

\subsubsection{The \hi-to-H$_2$ transition in diffuse environments}

The classical picture of the formation of molecules in \hi\ is that it requires gas that is both dense and well-shielded, meaning that the CNM is a necessary precursor to molecular cloud formation, and that molecules naturally form most readily within the densest and most shielded CNM structures. In this picture clumps of high molecular fraction are embedded within larger CNM envelopes; along any given sight-line, this corresponds to molecular gas distributed highly unevenly, confined to a limited number of host clouds \citep[e.g.][]{hafner2023,park2023}. 
However, \citet{liszt2000} and \citet{rybarczyk2022} have shown that, along some sight-lines, virtually all of the CNM hosts molecules, with many CNM components hosting apparently-diffuse molecular gas, in accordance with recent modeling (see Fig.~\ref{fig:P-n_HI-H2}). These observations, which used 3mm HCO$^+$ absorption as the molecular tracer, are complemented by the single-dish 18\,cm OH emission measurements of \citet{busch2021}, who find a tantalizingly similar result, with diffuse molecular gas present in all channels in which \hi\ emission is detected. 
Not only are the molecular gas properties associated with these diffuse structures systematically different than those of classical Giant Molecular Clouds (GMCs) --- the molecular gas is CO-dark, and the molecular fraction is low ($\lesssim10\%$) --- but so, too, are the properties of the atomic gas. Where thermal properties can be measured, the \hi\ associated with these very diffuse, partially-molecular gas detections has lower column densities, lower optical depths, and lower CNM fractions than \hi\ observed in the direction of GMCs \citep{rybarczyk2022}. There are early indications that these \hi\ structures have warmer average temperatures, but more work is needed to confirm this. 

Together, both observations and models suggest that H$_2$ may exist embedded in warmer, more diffuse atomic gas than previously understood. However, absorption measurements of the CNM have been obtained for only an extremely-limited sample of sightlines ($\lesssim10$) where  diffuse molecular gas has been detected. Such measurements are particularly important to constrain molecule formation and transport mechanisms. For example, a key mechanism implicated in the models of Figure \ref{fig:P-n_HI-H2} is the turbulent mixing of the CNM and WNM, which simultaneously (1) transports H$_2$, formed in colder and denser environments, into the more diffuse ISM (WNM, WIM), and (2) increases the fraction of \hi\ that exists as UNM \citep{valdivia2016, godard2023}. If this is the origin of very-diffuse H$_2$, we might expect to find the characteristic HCO$^+$ and OH signatures coincident with warmer \hi. Confirming this  would be essential to interpreting observations of diffuse molecular gas.

In synergy with existing millimeter and sub-millimeter facilities such as ALMA and NOEMA, SKA will revolutionize the studies of the \hi-to-H$_{2}$ transition by providing sensitive and matched \hi\ and molecular emission and absorption spectra toward many lines of sight, probing a wide range of environments. The SKA alone provides the possibility of matched OH and \hi\ absorption spectra against the same background sources, with the 18\,cm hyperfine lines of ground-state OH tracing both the dense and diffuse molecular component \citep[e.g.][]{allen2015, busch2021, dawson2022}. While the weakness of the OH lines means that only the brightest absorption sources are likely to reveal CO-dark H$_2$ \citep[e.g. moderate integrations with the VLA find almost entirely CO-bright gas,][]{rugel2025}, the SKA pathfinders are already revealing CO-dark OH towards multiple sources in the Galactic Plane \citep[GASKAP-OH,][]{dawson2024}. In addition, some 21-cm-bright sources will also be mm-bright. With dense spatial coverage of \hi\ absorption, sensitive HCO$^+$ follow-up observations will be feasible in some fraction of these directions, enabling us to map out both the atomic and (extremely diffuse) molecular gas.

\subsubsection{The \hi-to-H$_{2}$ transition in dense environments}

Examining the \hi-to-H$_{2}$ transition in dense environments where star formation occurs requires absorption measurements toward continuum sources that are distributed over a small area of the sky, and  
there have been only a handful of such studies so far \citep[e.g.][]{stanimirovic2014, denes2018, nguyen2019}.
\citet{park2023} is one such study, and showed that the following conditions are needed for the detection of CO(1--0) emission in the surroundings of local molecular clouds ($\lesssim 500$ pc): spin temperature $<$ 200 K, peak optical depth of the CNM $>$ 0.1, CNM fraction of $\sim$0.6, and $V$-band dust extinction $>$ 0.5 mag. 
Compared to the thresholds found by \citet{rybarczyk2022} for lines of sight at high Galactic latitudes of $|b| \gtrsim 10^{\circ}$, \citet{park2023} reported  slightly higher spin temperatures and $V$-band dust extinction, suggesting a possible difference in the conditions for the \hi-to-H$_{2}$ transition in diffuse and dense environments. 
However, this result remains to be confirmed with more \hi\ emission and absorption measurements, since the approaches of \citet{rybarczyk2022} and \citet{park2023} for constraining the \hi-to-H$_{2}$ transition are systematically different, e.g., \citet{rybarczyk2022} used HCO$^{+}$ absorption, while \citet{park2023} employed CO(1--0) emission. (As already discussed above, this allowed the former work to detect very diffuse, CO-dark H$_2$ alongside denser CO-bright components.) 

In addition to discrete point sources, the SKA will also enable us to spatially map the CNM and WNM across extended radio continuum sources in unprecedented detail, as recently demonstrated with the ASKAP telescope for the extreme star-forming region 30 Doradus \citep{park2026}. The utility of  simultaneous observations of 18-cm OH absorption also applies strongly here. Since the morphology of extended continuum is similar at 18 cm and 21 cm, the matched background source structure mitigates much of the ambiguity inherent in comparisons of \hi\ absorption with molecular emission lines such as CO(1--0).

In addition to discrete and extended continuum sources, \hi\ can be measured in \textit{self}-absorption against bright background \hi\ emission. Previous analyses of Galactic plane \hi\ survey data have revealed a population of \hi\ Self-Absorption (HISA) features, which vary in size, shape, and contrast against Galactic background \hi\ emission 
\citep[e.g.][]{gibson2000, gibson2005, kavars2005, mccluregriffiths2006, haud2013}. 
Comparative studies of HISA features and molecular tracers suggest that HISA structures are in the intermediate stage between diffuse atomic clouds and dense star-forming clumps \citep[e.g.][]{klaassen2005}. \hi\ Narrow Self-Absorption (HINSA) is similar in appearance to HISA, but has been typically extracted to match with molecular tracers (e.g., CO isotopologues, OH, and CH). As such, HINSA features tend to trace very cold \hi\ ($\lesssim$~10~K) that is located within the well-shielded portions of molecular clouds 
\citep[e.g.][]{li2003, goldsmith2005, tang2021}. 
Future SKA observations with high angular and velocity resolutions will enable us to extract HISA and HINSA features throughout the Galaxy. Follow-up multi-wavelength observations of those features (e.g., CO isotopologues, OH, CH, CRRLs) will then provide insights into the distribution of cold \hi\ in the context of the \hi-to-H$_{2}$ transition, as well as the chemical ages of molecular clouds in the Milky Way.

\subsection{Tiny Scale Atomic Structure}

\label{sec:TSAS}

As discussed in section~\ref{sec:structure}, the CNM has a multi-scale filamentary structure that extends from parsecs down to AU-scale, often termed Tiny Scale Atomic Structure (TSAS). Evidence comes from temporal fluctuations in H{\sc i} absorption spectra against fast-moving background pulsars \citep{frail1994,johnston2003,stanimirovic2010}, and very small scale spatial variations in absorption against resolved background sources \citep{deshpande2000a,brogan2005,lazio2009}. 
The implied size scales range from a few 1000 AU \citep{rybarczyk2020} down to a few AU \citep{stanimirovic2010} or even sub-AU scales \citep{liu2021,liu2025}, with optical depth fluctutions spanning two orders of magnitude and (on a point-by-point basis) showing no obvious correlation with lateral size \citep{stanimirovic2018}. 

The larger $\sim$0.01--0.1 pc ($\sim$2,000--20,000 AU) end of the TSAS distribution is relatively straightforward to measure \citep{rybarczyk2020}, and similarly straightforward to interpret, since these scales are similar to the width of CNM filaments (see section~\ref{sec:structure}). However, as we approach scales of $\lesssim 100$ AU, the experimental side becomes challenging, relying on VLBI imaging of complex, extended structure and temporal variations against pulsars (which are weak background sources). Claimed variations are often barely above the noise, and small errors in uncertainty propagation or failure to properly account for important systematics (e.g. H{\sc i} emission noise) could in principle render many detections non-significant. These concerns are exemplified by apparent discrepancies between VLBI studies --- which report copious $\sim$50 AU optical depth structure along sufficiently bright sightlines \citep[e.g.][]{brogan2005,lazio2009} --- and pulsar studies, in which non-detections appear to dominate  \citep[e.g.][]{johnston2003,minter2005,stanimirovic2010,liu2021}. The volume filling factor of TSAS is consequently unconstrained, with estimates ranging from between $10^{-6}$ to 0.01  \citep{stanimirovic2018}. Moreover, the physical interpretation --- particularly of optical depth variations on the smallest spatial scales --- is similarly uncertain, with proposed origins ranging from instability-triggered turbulent density enhancements \citep{stanimirovic2010}, projected ultra-cold edge-on sheets \citep{heiles1997}, to a superposition of turbulent structure requiring no enhanced densities at all \citep[e.g.][]{deshpande2000b}. 

The SKA will allow us to move from the realm of individual TSAS detections into the realm of TSAS statistics. In pulsar absorption experiments (which can in principle probe the smallest scales), all previous literature detections have been made against a mere 7 sources; with AA4, the viable source population is $\sim100$. For spatial variations, the source density  of sufficiently bright double-lobed background sources is likely to be $\sim1$ per square degree, opening up the possibility of piggybacking on deep extragalactic surveys to measure the statistics of spatial variations in local H{\sc i} absorption on $\sim100$--1000 AU scales. Excitingly, AA4 will permit high-sensitivity direct CNM imaging of scales as small as $\sim$1000 AU (via H{\sc i} self-absorption of targets like the Riegel-Crutcher Cloud, \citealt{mccluregriffiths2006}), allowing us to directly probe the morphological signatures of the instabilities implicated in structure formation. Together these experiments may usher in a step change in our understanding of the smallest neutral gas structures, facilitated by the SKA.


\section{\hi\ structure in the Solar neighborhood}

\label{sec:structure}

The 21\,cm absorption measurements presented in the previous section are the foundation of Galactic \hi\ physics, as they provide quantitative estimates of the cold gas spin temperature and optical depth. On the other hand, as powerful as this method is, it relies on the presence of bright background sources and therefore can provide only limited information about \hi\ morphology. This section describes the multi-scale and multi-phase structure of the atomic ISM as seen in \hi\, connects it with our knowledge of magnetic fields and the 3D distribution of neutral matter in the solar neighbourhood, and examines its relationship to the ionised phase. 


\subsection{Cold filamentary structure and its link to the magnetic field}
\label{sec:filaments}

\begin{figure*}
  \centering
  \includegraphics[width=0.505\linewidth]{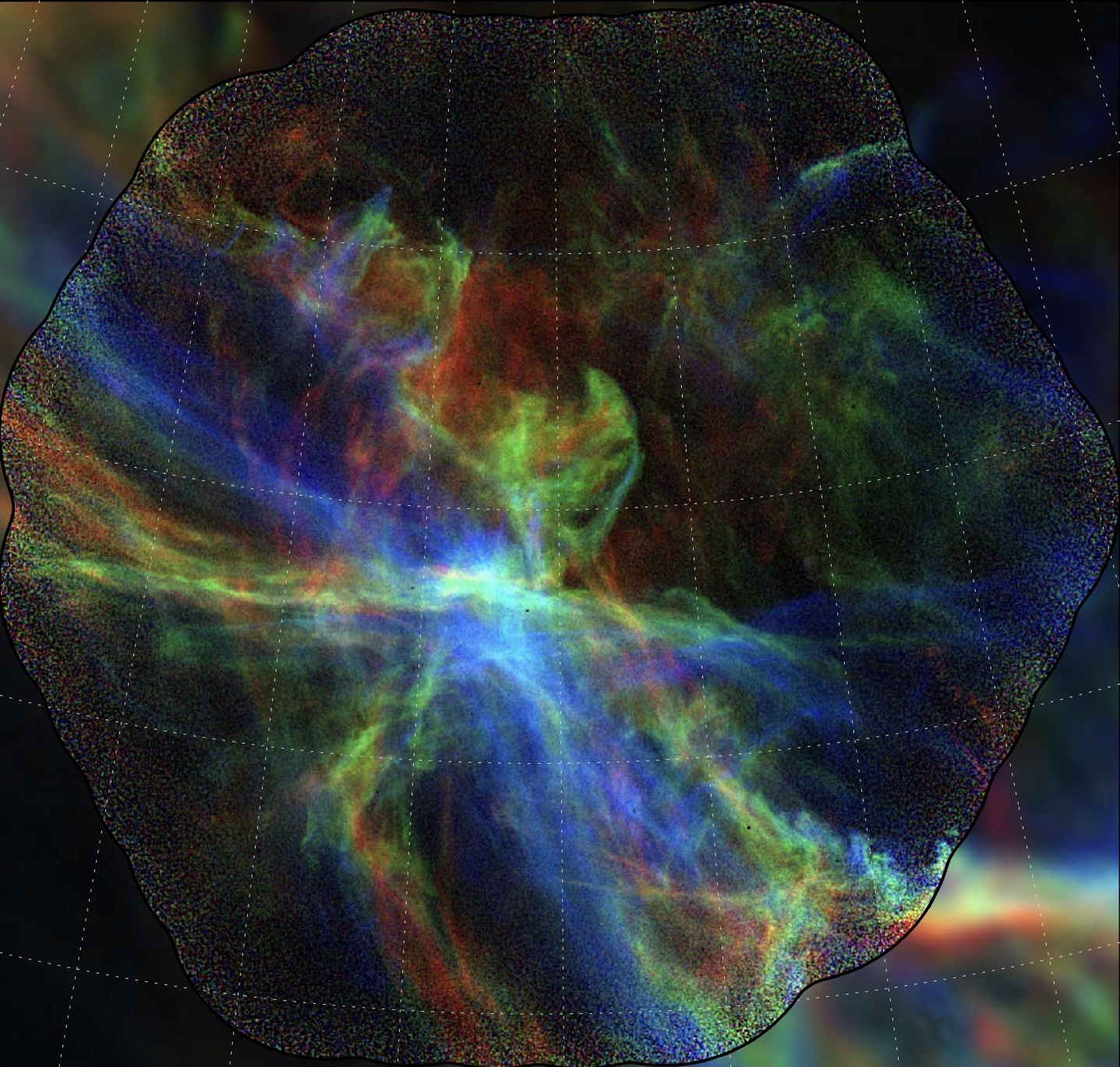}
  \includegraphics[width=0.48\linewidth]{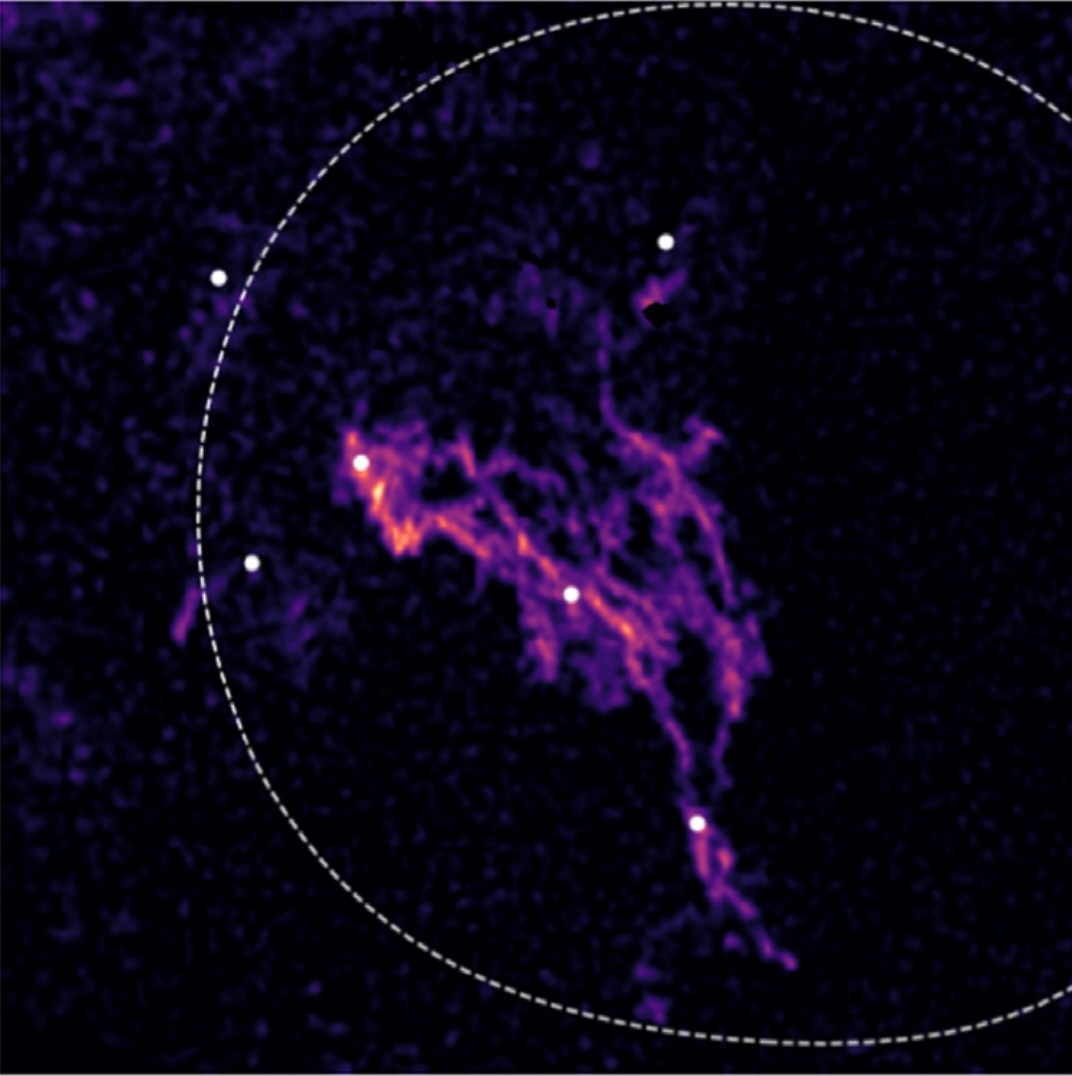}
  \caption{\label{fig:DHIGLS} Two illustrations of the \hi\ structure from DHIGLS, a 21\,cm survey of seven regions at high Galactic latitude, covering a total of 150 square degrees, using a combination of the DRAO interferometer and single dish, providing full scale sampling down to 1\,arcmin resolution \citep{blagrave2017}. {\bf Left: } The Galactic cirrus named Spider shown here as an RGB image combining three 21\,cm velocity channels at $7.59\,$km\,s$^{-1}$ (red), $5.12\,$km\,s$^{-1}$ (green), and $2.64\,$km\,s$^{-1}$ (blue). For each color the identical intensity range was used, from $0$\,K to $40$\,K. The field size is $7.6^\circ \times 7.6^\circ$  \citep{blagrave2017}. {\bf Right: } A $2.5^\circ \times 2.5^\circ$ area of Complex\,C, a Milky Way halo cloud. Here is shown the CNM column density after phase decomposition of the 21\,cm emission data (range is $0 \leq N_{\text{\hi}}  \leq 7\times 10^{19}\,$cm$^{-2}$). See \citet{marchal2021} for details. }
\end{figure*}

\subsubsection{Alignment between \hi, dust polarization and Faraday structures}

The advent of sensitive, large-area \hi\ surveys has facilitated a deeper understanding of the physics that drives the evolution of gas in galaxies. Much of the \hi\ emission in the Solar neighborhood and throughout the Galactic disk is found in high-aspect ratio structures generically known as ``filaments'' \citep[Fig.~\ref{fig:DHIGLS} and][]{heiles1984,mccluregriffiths2006, clark2014,soler2022, Hacar:2023}. Some of these H{\sc i} filaments extend from the midplane like ``worms'' crawling away from the Galactic plane, most likely as the result of the bursting of \hi\ shells and supershells toward the inner Galaxy \citep{pidopryhora2007,koo1992,soler2022}. Interferometric observations of these filamentary structures indicate that they can extend across more than one kiloparsec and host up to 10$^{5}$\,M$_{\rm \odot}$     
\citep{soler2020,wang2020,syed2022}, thus representing important gas reservoirs and crucial tracers of large-scale Galactic dynamics. The key physics that remains poorly understood is their connection to the molecular gas structures at smaller scales, but our understanding is limited by the angular resolution and sensitivity of the existing observations. 

\begin{figure*}
  \centering
  \includegraphics[width=0.48\linewidth]{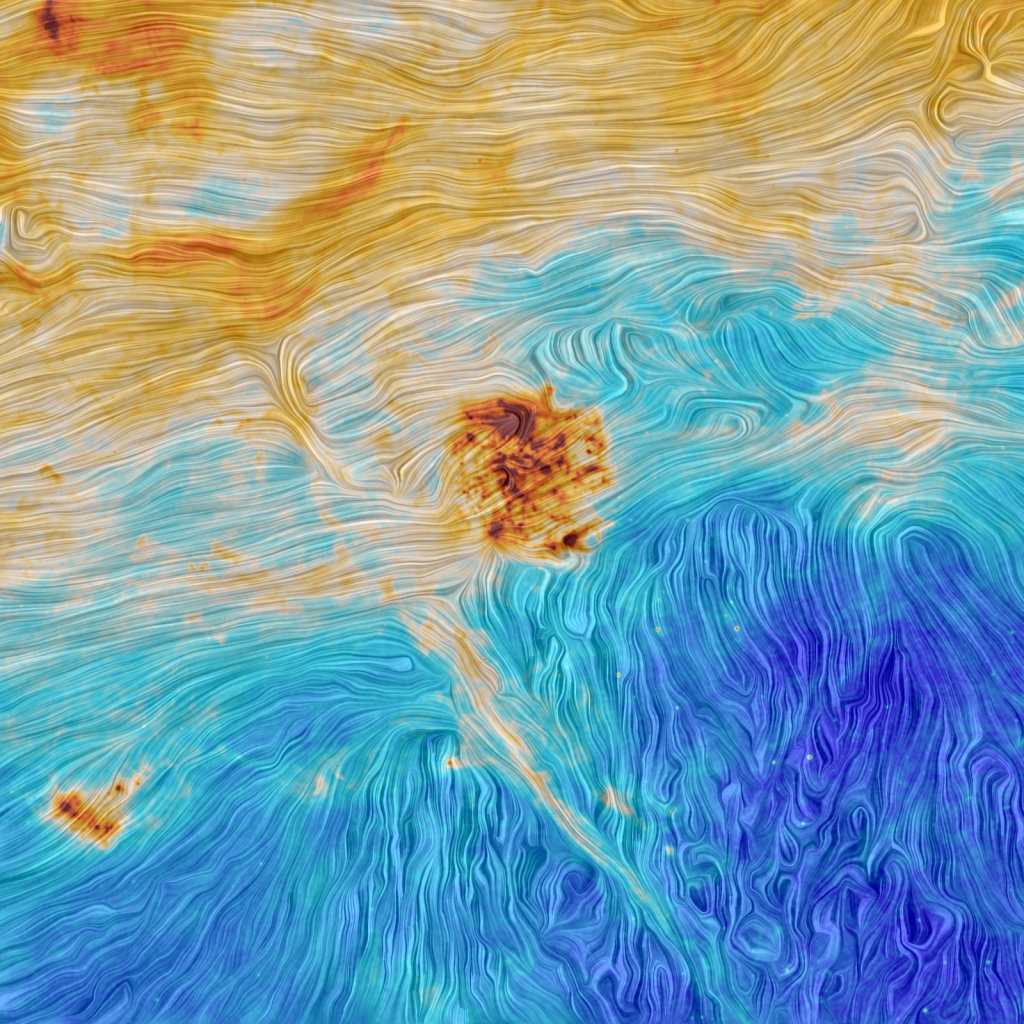}
    \includegraphics[width=0.48\linewidth]{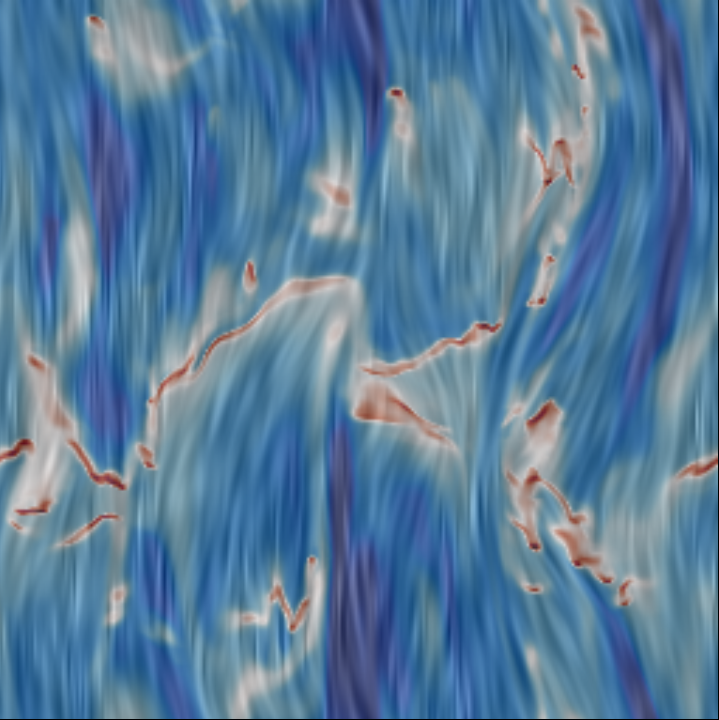}
  \caption{{\bf Left:} The Magellanic cloud area observed with Planck; background color indicates ISM column density deduced from dust emission, while the Line Integral Convolution (LIC) pattern shows the projected orientation of the magnetic field estimated from 353\,GHz dust polarization. Below the LMC (bright galaxy in the centre of the image) is a 100\,pc long \hi\ filament fully aligned with $B$. {\bf Right:} density slice of a 3D ``turbulence-in-a-box'' MHD simulation with cooling from \citet{berat2026}. In such simulations the alignment between small scale cold structures (in red) and the $B$ field (LIC pattern) is often not as strong as what is inferred from observations, indicating an incomplete understanding of the multi-phase dynamics.}\label{fig:Planck_MC}
\end{figure*}

The filamentary \hi\ emission of the Milky Way, shows elongated structures that are strikingly aligned with the interstellar magnetic field as traced by Planck dust polarization measurements \citep[][and Fig.~\ref{fig:Planck_MC}]{clark2014, clark2015, kim23}. These structures are primarily associated with the cold neutral medium \citep{kalberla2016, Clark:2019, peek2019b, kalberla2025, putman26}. High-resolution observations of these filaments have opened windows into the phase structure of the ISM \citep{marchal2021a,lei2023}, the three-dimensional structure of the interstellar magnetic field \citep{Clark:2018, Clark:2019, Pelgrims:2021}, the nature of the polarized dust foreground to the cosmic microwave background \citep{Clark:2021, Cukierman:2023, Ade:2023, Halal:2024}, and the link between the phase structure and magnetic properties of the ISM \citep{Bracco2020, Campbell:2022, Lei:2024, Nowotka:2025}. 
The morphology of diffuse H{\sc i} emission is thus a valuable probe of the physics of the diffuse universe. Complementary theoretical effort has focused on understanding the combined effects of thermal instability, the turbulent velocity field, and the anisotropy introduced by the interstellar magnetic field \citep{inoue2016,villagran2018,hennebelle2019}.
Both the cold and warm ``neutral'' \hi\ phases have an ionization fraction sufficient to couple the gas to the magnetic field, which exerts an anisotropic influence on the gas motion, but the strong alignment between density and magnetic field inferred from observations remains to be understood \citep[Fig.~\ref{fig:Planck_MC} and][]{berat2026}.

Observationally our current understanding of the anisotropy of the cold \hi\, and its link with the magnetic field was enabled by the combination of the all-sky Planck data \citep{planck_collaboration2015f}, LOFAR low-frequency radio-polarization data \citep[e.g.][]{Bracco2020,berat2026} and the sensitivity and angular resolution afforded by the GALFA-\hi\ survey \citep{peek2018}, but no low-frequency data comparable to LOFAR exist in the Southern hemisphere. A similar survey with SKA would revolutionize the study of \hi, providing an unprecedented window onto the evolution of gas from the most diffuse ISM phases to star-forming regions, and the impact of stellar feedback on the interstellar environment. 

\subsubsection{Zeeman splitting of the 21\,cm line}

\label{sec:Zeeman}

A critical aspect to understand the influence of the magnetic field is to measure its \textit{strength}. One of the most direct diagnostics in \hi is Zeeman splitting of the $21\,$cm line that provide estimates of the line-of-sight magnetic field strength ($B_{\mathrm{LOS}}$). A systematic survey of the Zeeman effect in \hi\ absorption towards the brightest continuum sources was conducted with the Arecibo Observatory \citep{heiles2003}, producing $B_{\mathrm{LOS}}$ measurements for 69 distinct velocity components, with a median of 6 $\mu$G. Based on that dataset, \citet{Heiles:2005} found an approximate equipartition between the magnetic and turbulent energy densities, both dominant over thermal energy density.
Progress since the Arecibo survey has been limited by instrument sensitivity and the sparsity of bright background sources. Over the years, there have been a few interferometric measurements of the Zeeman effect in \hi\ absorption \citep[e.g., toward Orion’s Veil;][]{Troland:2016}, and several of the Arecibo measurements have recently been independently confirmed with FAST \citep{Nowotka:2025}. The Zeeman effect has also recently been detected in \hi\ narrow self-absorption (HINSA) towards a pre-stellar core \citep{Ching:2022}. 

A large-scale \hi\ Zeeman campaign with the SKA would address a number of urgent questions. How are magnetic fields and \hi\ structures linked, especially at higher resolution? How does feedback affect the magnetic field strength throughout the ISM? Do magnetic fields accelerate or inhibit the formation of star-forming clouds, especially across the earliest stages of their coalescence? How does the gas density scale with magnetic field strength, and can we understand this in terms of fundamental physical processes? These questions would be powerfully addressed by a large statistical sample of Zeeman measurements, in combination with a sensitive survey of \hi\ emission.
SKA will make \hi\ Zeeman measurements in the South for the first time with excellent sensitivity, thus enabling measurements toward fainter sources and/or shallower absorbers. The SKA's combination of sensitivity and spatial resolution will also be ideal for mapping magnetic fields in regions with extended \hi\ self-absorption, e.g., the Riegel-Crutcher cloud \citep{mccluregriffiths2006}.

\subsection{The CNM-WNM relation from emission data}

\label{sec:phases-from-emission}

As mentionned in earlier sections, theory, and 21\,cm absorption and emission measurements indicate that the CNM is distributed in dense, filamentary structures, in contrast to the more diffuse, volume filling, WNM \cite[e.g.][]{mccluregriffiths2006}. Due to their very different kinetic temperatures ($\sim 80\,$ and $\sim 6000\,$K), the CNM and WNM have significantly different thermal line-widths ($\sim 0.8$ and $\sim 7.0$\,km\,s$^{-1}$). Those physical differences leave different imprints in the \hi\ hyperspectral data\footnote{position-position-velocity, PPV, data cubes}, both spatially and spectrally.

Recently, several methods have exploited this rich information to segment phases out of the emission data themselves.
For instance, \citet{marchal2019} developed a spatially coherent decomposition algorithm that models data cubes as sums of Gaussian components, while maintaining physical interpretability through the enforced smoothness of the solution.
Leveraging the fast emergence of Machine Learning methods, \citet{murray2020} exploited the spectral information of 21\,cm spectra to train a Convolutional Neural Network (CNN) that infers the spin temperature and column density correction factor for optically thick \hi, with application demonstrated on the GALFA-\hi\ survey at high Galactic latitudes.  
\citet{Nguyen:2025} improved on this pioneering work using a neural network combining Convolutional and Transformer architectures with Positional encodings that shows a 10\% increase in accuracy, stability, and convergence speed compared to deep CNNs.
Finally, \citet{Lei:2025} developed a morphology-based phase decomposition model that combines scattering transform (ST) statistics with a variational autoencoder (VAE). This data-driven framework is capable of learning statistical models of the morphological features of different \hi\ phases directly from data without relying on simulation training. Applying the ST+VAE model to GALFA-\hi\ data, \citet{Lei:2025} produced denoised CNM and WNM maps of the diffuse sky at 4' and 3 km\,s$^{-1}$ angular and spectral resolution.

\begin{figure*}
  \centering
  \includegraphics[width=\linewidth]{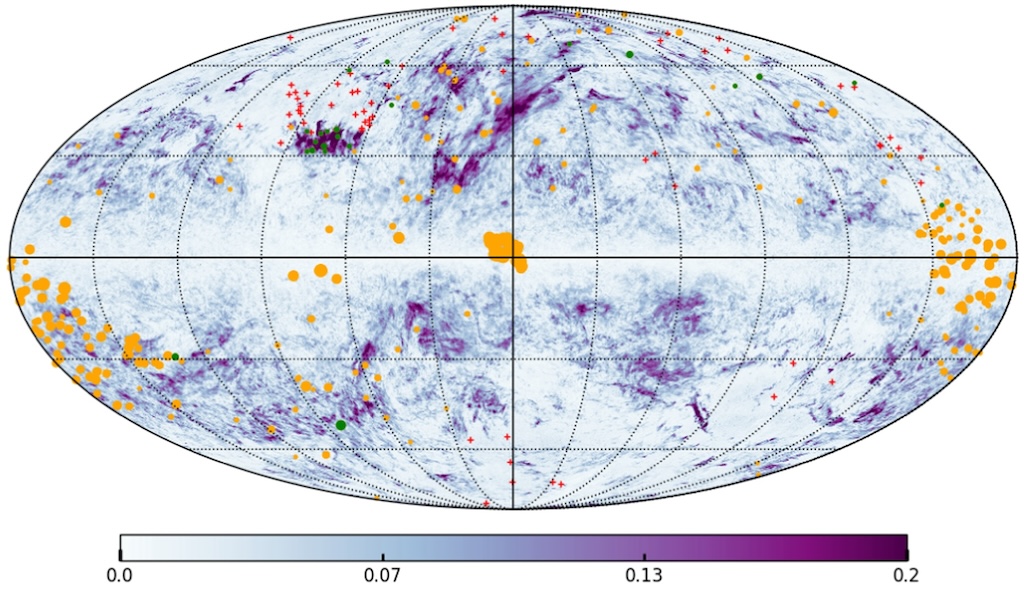}
  \caption{From \cite{Marchal:2024}. Mollweide projection centered on the Galactic center of the lower limit on the cold gas mass fraction of the entire HI4PI survey in the velocity range -90 < $v$ < 90\,km\,s$^{-1}$. The red crosses and orange+green circles show the positions of non-detections and detections, respectively, in the compilation of \hi\ absorption spectra from \citep{McClure-Griffiths2023}.}
  \label{fig:marchal2024}
\end{figure*}

Being able to extract the morphological structure of \hi\ components and to sort them according to their linewidths is opening the exploration space drastically. For instance, the comparison of CNM extracted from emission data with dust column densities indicates that cold \hi\ structures contain a significant amount of CO-dark H$_2$ \citep[e.g.][]{kalberla2020}. Such segmentation methods applied to 21\,cm data of relatively isolated systems in position-position-velocity data have allowed the study of the multi-phase structure of the high-velocity cloud complex C \citep{marchal2021}, and revealed the spatial distribution of cold gas in the North Celestial Pole Loop \citep{Taank:2022,Marchal:2023}, the LLIV-Arch IV \citep{vujeva2023}, clouds in the halo of the Small Magellanic Cloud \citep[][and section~\ref{sec:halos}]{bucklandwillis2025}, and the Corona Australis molecular cloud \citep{Bracco:2020}. These studies are revealing a CNM that is very structured at small scales, often related to dynamical events \citep[compression, dynamical instabilities, see][]{marchal2021}, even showing areas with 100\% of the \hi\ in the cold phase (see section~\ref{fig:SMC-halo}). 

Using the spectral information of 21\,cm data cubes, \citet{Marchal:2024} was able to separate broad and narrow Fourier modes of individual spectra with low computing time and memory. This method produced a map of a lower limit of cold gas mass fraction over the whole high Galactic latitude sky (see Figure~\ref{fig:marchal2024}). Like \citet{murray2020}, it revealed a high level of variability of the CNM abundance within the Local Bubble volume\footnote{Here we refer to the Local Bubble as the volume defined by recent three-dimensional reconstructions of dust made using Gaia data \citep[e.g.][]{lallement2019,edenhofer2024}}, on a large range of scales. This is revealing new locations of cold gas formation that would have remained unknown if the exploration was limited to CO-bright areas \citep[similar to diffuse H$_2$ fluorescence detected from previously unknown molecular clouds by][]{burkhart2025}. 

The CNM, with its contrasted and filamentary structure, and its narrow line-width, is very striking in 21\,cm data of the Milky Way. On the other hand, more than half of the emission comes from the WNM, a more diffuse and spectrally broad component. Using the same data segmentation techniques to isolate the CNM \citet{marchal2021} was able to extract the WNM on a part of the Local Bubble. They found that the WNM occupies about $50\%$ of the volume in contrast with the CNM clouds that have a volume filling factor of about $1\%$. This diffuse phase of the \hi\ is also more extended with a scale height about twice that of the CNM \citep{dickey1990,kalberla2009}. The WNM has has the properties of a subsonic turbulent flow \citep[Mach $\sim0.9$ at the largest scale observed of $130\,$pc by][]{marchal2021}. Interestingly the WNM seems to have a level of turbulence similar to that observed in the densest phase (CNM, molecular clouds) with turbulent velocity dispersion at a scale of $1\,$pc of $0.76\,$km\,s$^{-1}$ \citep{hennebelle2012,miville-deschenes2017}. This widespread diffuse neutral phase is the matrix out of which the cloud condensation process occurs. 

\subsection{The relationship of the \hi\ with the ionized phase}

As discussed above, \hi\ is partially ionized, albeit at a low level. There is now growing evidence that the WNM is more connected than previously thought to more highly ionized diffuse ISM phases -- 
the diffuse photoionized warm ionized medium, WIM, (sometimes called the diffuse ionized gas, DIG) and the ``warm partially-ionized medium'' (WPIM, $n_e / n_H \sim 1/2$, \citealt{heileshaverkorn2012}).
While the WIM is fully ionized \citep[$n_e/n_H \gtrsim 0.9$;][]{reynolds1998}, the WNM remains partially ionized by cosmic rays at an ionization fraction of $n_e / n_H \sim 10^{-3}$. In addition, supernovae shock waves propagating in the diffuse ISM could also contribute significantly to the ionization of the gas. Depending on the local conditions (strength of the ambient radiation field, presence of shocks) these diffuse phases will have different morphologies and volume filling fractions \citep[see][for a recent application to the Local Bubble]{mccallum2025:halphasky}. 

The ionized and partially-ionized phases are revealed by different tracers. Higher-density phases are directly revealed by emission lines from recombining gas. H$\alpha$ in the optical is by far the most sensitive (emission measures $\EM = \int n_e^2 ds \sim 0.1 \pccmsix$) and reveals a pervasive WIM \citep{haffner2009} but is subject to extinction, and the widespread, kinematically-resolved survey with the Wisconsin H-Alpha Mapper (WHAM) is limited to $1^\circ$ angular resolution \citep{haffner2003}. Wide-area radio recombination line (RRL) surveys are not subject to extinction and can have angular resolution close to $1'$ at $5 \GHz$ with the GBT, but to date only reach $\EM \sim 1100 \pccmsix$ \citep{anderson2021}. With a wide bandwidth to stack dozens of RRLs in Band 2, or fewer lines with a higher line-to-continuum ratio at higher frequencies, SKA promises to bring RRL studies close to the EM sensitivity of H$\alpha$ observations but with sub-arcminute resolution and without extinction.

Meanwhile, Faraday rotation probes magneto-ionized gas, sensitive to the Faraday depth $\phi \equiv 0.81 \int n_e \vec{B} \cdot d\hat{s}$. Rotation measure (RM) grid observations of point sources \citep{vaneck2021,hutschenreuter2022,gaensler2025} tend to provide complementary information to Faraday rotation of diffuse synchrotron emission. At higher frequencies ($\sim \GHz$), Faraday rotation observations are sensitive to Faraday depth differences of $\sim$ tens of \radmsq, corresponding to $n_e \sim 10 \cucm$ for typical line of sight magnetic field strengths $B_{||} \sim 5 \uG$; here Faraday rotation likely probes the WIM. Faraday rotation observations can reveal low electron column gas that is not evident in any other tracers, for example in the wake of a high-velocity cloud \citep{hill2013,betti2019} or in an isolated cloud of unknown origin \citep{mohammed2024}. At LOFAR/SKA-Low frequencies, synchrotron emission passing through the WIM is depolarized such that Faraday rotation observations are sensitive only to the WNM \citep{vaneck2017,Bracco2020}; there is also evidence that pulsar RMs at $1.4 \GHz$ are dominated by the partially-ionized WNM \citep{foster2013}.

Complementary \hi, RRL, and Faraday rotation are needed to understand the relative distributions of the neutral, partially-ionized, and fully-ionized gas. These observations promise to enable detailed studies of the ionization mechanism of the WIM \citep{domgorgen1994,wood2000,haffner2009,wood2010,vandenbroucke2018}, further enabled by the use of modern 3D dust maps (see Section~\ref{sec:3D}) coupled to photoionization by realistically-placed O stars \citep{mccallum2025:halphasky}. Bringing the angular resolution of RRL and diffuse polarization surveys close to that of \hi\ will be crucial for disentangling the extent to which each of these observations probe physically-related gas. To be sensitive to the large-scale structure of the Milky Way in diffuse emission, Faraday rotation surveys should cover $\sim 300$--$1800 \MHz$, essentially bands 1 and 2. Meanwhile diffuse polarization observations with SKA-Low likely probe primarily the WNM. Therefore jointly-designed surveys across this full band will be sensitive to the physics establishing the multi-phase nature of the Milky Way ISM.

\subsection{3D Structure of Neutral Gas in the Solar Neighborhood}
\label{sec:3D}

All-sky single-dish surveys like HI4PI \citep{hi4pi} have revealed 21\,cm emission everywhere, with structure at all scales; there is indeed not a single line of sight without Galactic \hi. Away from the Galacic plane, most of the sky ($>80$\%) samples the Solar neighborhood, the gas located in the Local Bubble, 200--300 pc away from the Sun, as well as the \hi\ in the Milky Way halo. This volume around the Sun is a local laboratory where the evolution of interstellar matter can be studied in details on scales from milli-pcs to kpcs. All-sky \hi\ data is now part of the current effort to reconstruct in 3D as many physical properties of the interstellar medium as possible. 

Over the past five years, 3D dust mapping has revolutionized our ability to chart the three-dimensional spatial structure of neutral gas in the solar neighborhood. By combining Gaia parallaxes with multiband photometry and low-resolution spectroscopy, recent 3D dust maps have achieved parsec-scale spatial resolution out to roughly 1–2 kpc from the Sun—an improvement of nearly two orders of magnitude over pre-Gaia work \citep{edenhofer2024,Leike2020,LallementVergely2022}. This resolution is comparable to that achieved for nearby, face-on galaxies observed with JWST out to a few Mpc \citep{LeeSandstrom2023}, allowing the local Milky Way to be studied through the same lens normally reserved for extragalactic systems.

Because dust and neutral hydrogen are tightly correlated, these 3D dust maps provide an effective framework for reconstructing the distribution of \hi\ in three spatial dimensions. The densities of the cold and warm neutral media fall squarely within the regime where 3D dust mapping is most sensitive ($\simeq$0.1–100 cm$^{-3}$). Consequently, 3D dust reconstructions naturally capture the \hi-dominated gas in the local volume, revealing a complex network of molecular clouds, cavities, and large-scale spiral structure. For example, we now have detailed 3D spatial models and physical constraints (e.g., masses, densities, and thicknesses) for feedback-driven superbubbles (the Local Bubble and IRAS Vela Shell; \citealt{ONeill2024,GaoZucker2025}), molecular clouds \citep{ZuckerGoodman2021}, and large-scale spiral features \citep{AlvesZucker2020,KuhnBenjamin2021} out to a few kiloparsecs in distance.

We can also move beyond purely spatial reconstructions into spatial–dynamical reconstructions that trace the flow of neutral gas near the Sun. The combination of modern Gaia-enabled 3D dust maps, high-resolution stellar spectroscopy, and wide-field \hi\ 21\,cm surveys has given rise to a new, data-driven framework known as kinetic tomography \citep{ZasowskiFinkbeiner2019}. Kinetic tomography combines distance-resolved tracers of ISM density with spectral-line velocity information to infer the distance–velocity structure of neutral gas. By morphologically matching 3D dust structures to velocity-resolved \hi\ emission, this technique has enabled the reconstruction of four-dimensional (3D space + 1D velocity) maps of the local interstellar medium \citep{tchernyshyov2017,soler2025,SolerZucker2023}. Complementary point-based approaches use interstellar absorption features (e.g., diffuse interstellar bands or NaI D; \citealt{SaydjariUzsoy2023,LallementWelsh2003}) in high-resolution stellar spectra to directly measure integrated velocity and column density toward stars of known distance, which can similarly be decomposed into 3D maps of neutral gas flows \citep{tchernyshyov2018,DucheneHottier2023}.

This coupling of spatial and kinematic information transforms traditional views of \hi\ emission into dynamic reconstructions that capture how gas moves through the solar neighborhood—its expansion within superbubbles, condensation into clouds, and eventual return to the disk through large-scale fountain flows. In future, such analyses will provide a powerful means to quantify the pressure balance and turbulent energy of the neutral ISM, offering a direct, data-driven view of how \hi\ circulates, dissipates energy, and couples to feedback on Galactic scales. The enhanced sensitivity, angular resolution, and sky coverage of the SKA will provide the critical foundation for constructing higher-dimensional models of the interstellar medium, extending these dynamic reconstructions from the solar neighborhood to the full Galactic disk and halo.


\section{The \hi\ on Galactic scales}
\label{sec:multi-phase-MW}

\subsection{The Milky Way disk}

21\,cm emission data has been used for decades to unveil the large scale structure of the Milky Way disk. We know that \hi\ extends as far as $60\,$kpc from the Galactic center \citep{kalberla2009}, much further than the stellar and the molecular components \citep{miville-deschenes2017}. The \hi\ disk is also warped and flared, with its scale-height increasing from $\sim 100\,$pc in the inner Galaxy to 2--3\,kpc at $R_g>30\,$kpc. 
Because the vertical structure of the disk is set by the star formation cycle (e.g., stellar feedback and supernovae, the magnetic field, gas pressure, turbulence, cosmic rays) and by disk dynamical processes and interactions \citep[e.g.,][]{hunter1969,weinberg2006}, constraints on the $z$ distribution of multiphase \hi\ provide key tests for ISM models and simulations \citep[e.g.,][]{hill2018,hopkins2018}.
While the $z$ distribution of the total \hi\ has been well characterized through previous emission observations, constraints on the $z$ distribution of the CNM are relatively poor, largely due to the paucity of adequate \hi\ absorption data. Current literature values are discrepant regarding the thickness of the CNM layer as a function of Galacto-centric radius \citep{dickey2009,dickey2022,rybarczyk2024}.

What $21\,$cm absorption measurements obtained so far in the Milky Way disk have revealed is that the CNM is present as far as $R_g=40\,$kpc \citep{strasser2007,dickey2009,dickey2022}. Not only does the CNM disk extend much further than the molecular one, the cold \hi\ fraction, $f_{\rm CNM}$, seems to be almost constant at $15$--$20\%$. 
This result remains unexplained. Indeed CNM formation is though to be partly driven by compression and energy injection provided by stellar feedback, but the star formation activity is very low in the outer Galaxy. In addition, at $R_g=35\,$kpc the radiation field intensity and metallicity are respectively 100 and 10 lower than in the Solar neighborhood. Theory \citep{wolfire2003} and numerical simulations \citep{smith2023} predict that the CNM fraction should be significantly lower in these conditions. 

\begin{figure*}
  \centering
  \includegraphics[width=0.534\linewidth]{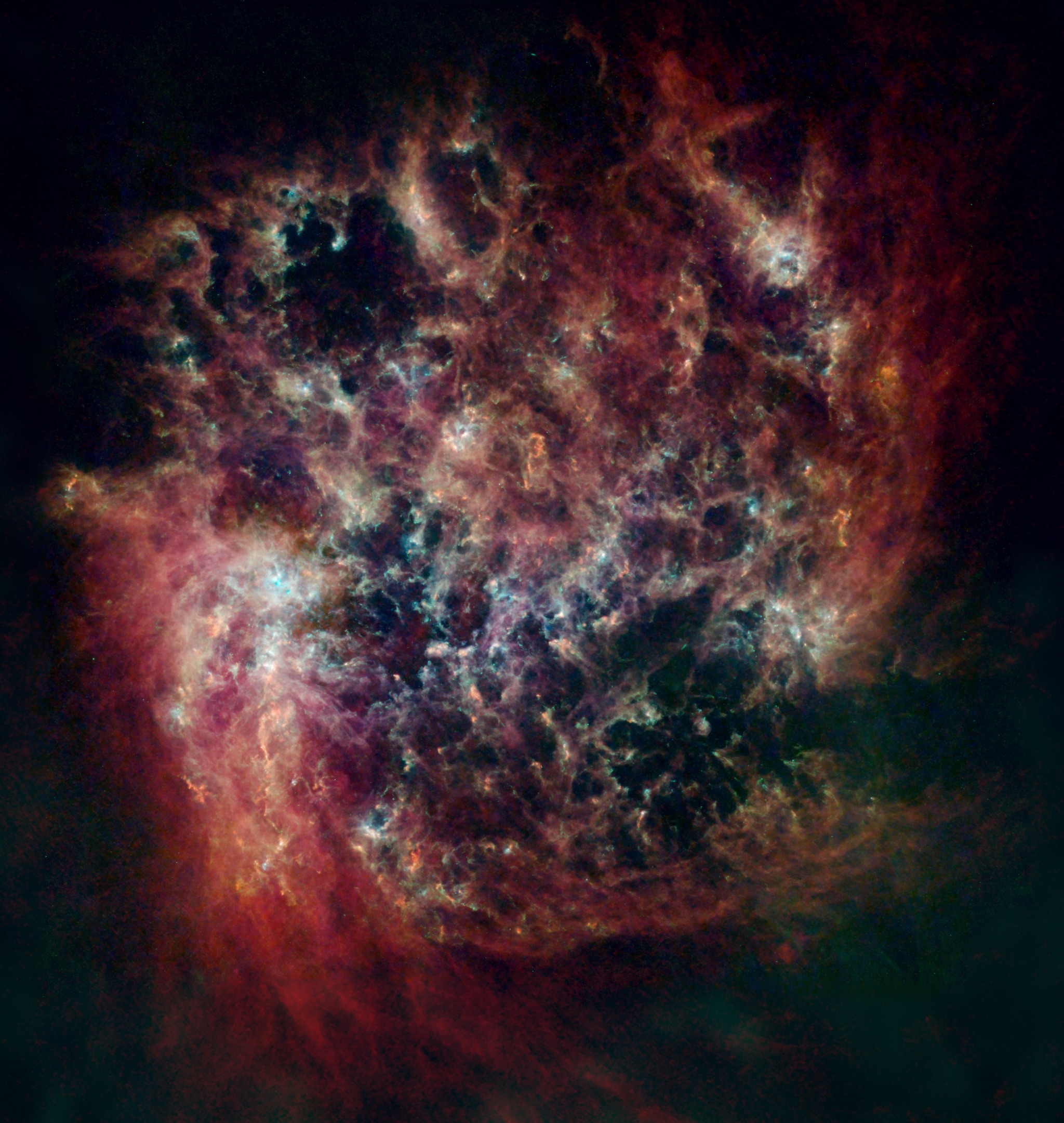}
    \includegraphics[width=0.45\linewidth]{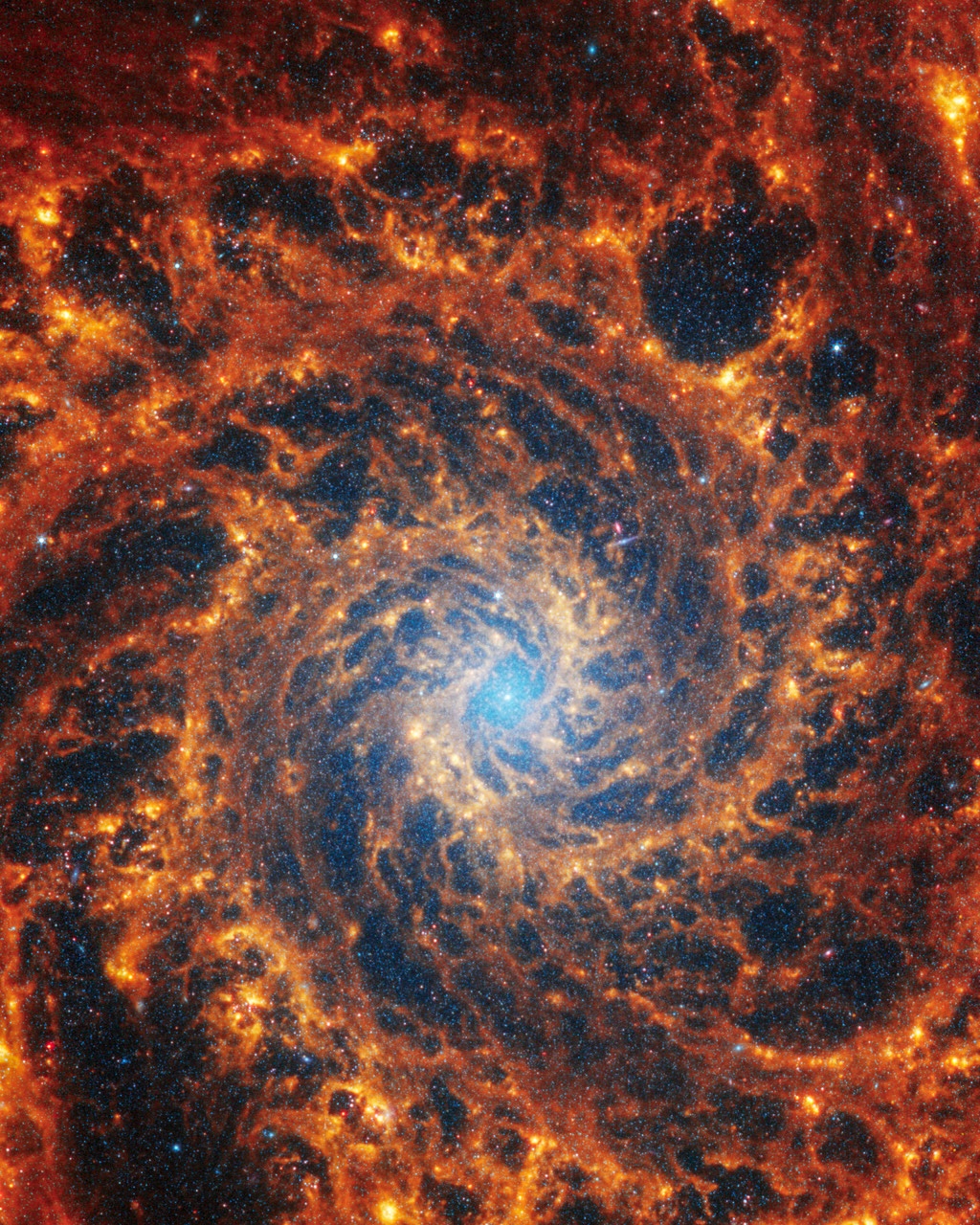}
  \caption{Large scale structures of two nearby galaxies. {\bf Left:} the LMC where red is \hi\ (Parkes and ATCA data), green is cold dust (Planck) and blue is warmer dust (IRAS) - credit Christopher Clark. {\bf Right:} Central zone of NGC~628 imaged in the mid-infrared (PAH emission) by JWST. }
  \label{fig:LMC_and_NGC628}
\end{figure*}

\subsection{Multi-phase \hi\ in nearby galaxies}

\label{sec:external-galaxies}

Nearby galaxies (the Magellanic Clouds and beyond) provide two fundamental opportunities to study the \hi: (1) they free us from the confusion of looking at the Milky Way from within to understand its large-scale properties, and (2) they provide a way to expand the study of \hi\ physics and the gas condensation process to a much larger range of physical conditions than present in the Solar neighborhood, including different metallicities \citep{stanimirovic2024}. 

\paragraph{\hi\ emission:}
A wealth of information on galaxy evolution and the star formation cycle is revealed through the large-scale structure and physical properties of the \hi\ inferred from 21\,cm emission data of nearby galaxies \citep{braun1997,kim1998,stanimirovic1999,thilker2004,walter2008,de_blok2018,koch2018}. For example, \cite{elmegreen2001,padoan2001} estimated the thickness of the LMC ($\sim$100--180\,pc) by studying the statistical properties of \hi\ emission.
The LMC, like other face-on galaxies, shows holes or voids in the matter distribution (Fig.~\ref{fig:LMC_and_NGC628}). Voids with sizes smaller or comparable to the scale-height of the disk may be attributed to stellar feedback (\hi\ supershells), while larger ones may be better be explained as the result of large-scale gravitational instabilities in the disk \citep[e.g.][]{wada2000,dib2005}. However, the distinction is by no means clear either observationally or theoretically. 
Here, \hi\ observations offer potentially valuable information, because stellar feedback and disk gravitational instabilities induce different pressure excursions (stellar feedback is more compressive). This in turn has an impact on CNM formation, which is more effective in compressive flows \citep{saury2014}. 
As the 21\,cm emission contains information about the kinematics and thermal state of the gas ($T_k$, $f_{\rm CNM}$, Mach number) \hi\ could bring important information on the nature of the origin of disk structures, and hence also on large-scale turbulence driving in disk galaxies, which remains a fundamental open question \citep{colman2025}.

Relevant to this study are three recent / ongoing large observing programs of nearby galaxies :
\begin{enumerate}
\item GASKAP-\hi\ \citep{pingel2022} is mapping the Magellanic Clouds with ASKAP providing 21\,cm emission data cubes of our closest neighbours at an angular resolution of 30" and a velocity resolution of 0.98 km s$^{-1}$;
\item The Local Group L-Band Survey \citep[LGLBS,][]{koch2025} is an extra-large project ($>1700$~hours at 0.4 km\,s$^{-1}$ resolution) undertaken with the VLA and dedicated to mapping several prominent members of the Local Group---including the Milky Way analogue, M31;  
\item MHONGOOSE \citep{deblok2024} is a large program surveying 30 nearby galaxies in the southern hemisphere with MeerKAT.
  \end{enumerate}
These three surveys allow imaging of nearby galaxies at resolutions $\leq 30$", which translates to physical resolutions as small as $\sim 10\,$pc. The selected targets span a wide range of astrophysical environments (e.g., star formation rate, metallicity, and gas-to-dust ratios)---all of which influence the \hi\ condensation process and the local star-forming efficiencies and rates.
These surveys also made the choice to observe at high  spectral resolution ($\leq 1.4\,$km\,s$^{-1}$), a key requirement in mapping the CNM and the small scale details of the gas dynamics. Such a strategy should be pursued with SKA observations of nearby systems.

Some attempts at examining the spatial distribution of the CNM in galaxies have been made in the past \citep[e.g.,][]{warren2012,patra2016,saikia2020,biswas2022,Park_HJ:2025}. All these studies identified narrow \hi\ lines ($\sigma<6\,$km\,s$^{-1}$), often found in localized areas, away from high surface density and/or star forming regions. This might be an observational bias due to the difficulty in identifying narrow lines in confused sight-lines, but an interesting outcome is that CNM is detected at large distances from the center of galaxies, similarly to what is found in the outer disk of the Milky Way \citep[see \S~\ref{sec:multi-phase-MW} and ][]{dickey2022}.

\paragraph{\hi\ absorption:}
Absorption studies are more challenging for extra-galactic systems. Unlike for the Milky Way for which there is always Galactic \hi\ along the line-of-sight towards background radio sources,  the small solid angle subtended by external galaxies lowers the probability of intersecting sufficiently bright background sources. The exception is the Magellanic System that spans several hundreds of square degrees on the sky. 
In both the Large and Small Magellanic Clouds, targeted observations with the Australian Telescope Compact Array (ATCA) have detected \hi\ absorption against a few 10s of fields, revealing spin temperatures that appear lower than typical values found in the Milky Way  \citep{dickey1994,marxzimmer2000, dickey2000,jameson2019}. \hi\ absorption was also detected in the Magellanic Stream for the first time by \citet{dempsey2020}, giving $T_s=68\pm20$\,K.

\begin{figure*}
  \centering
  \includegraphics[width=\linewidth]{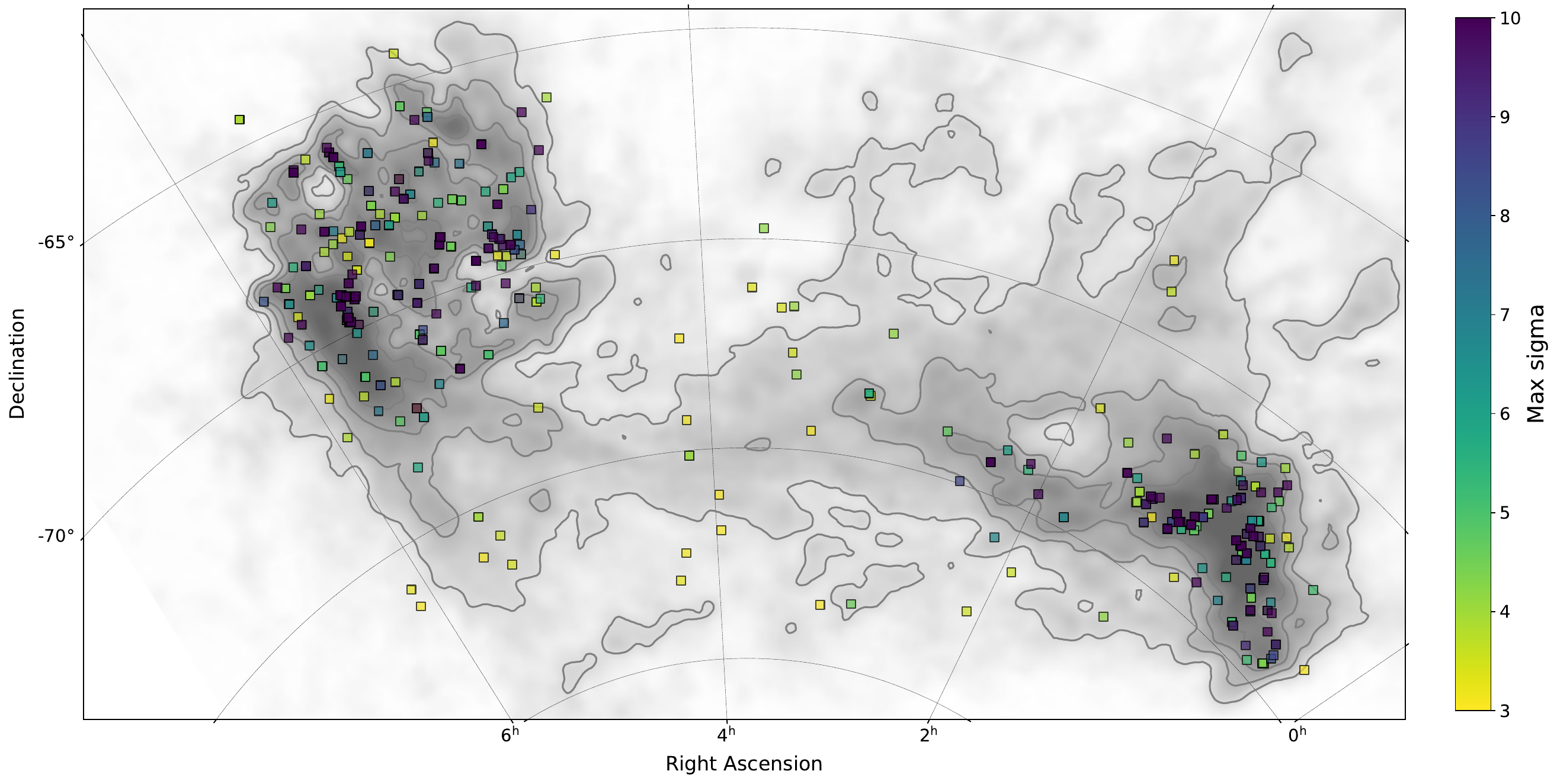}
  \caption{From \citet{dempsey2026} Plot of candidate absorption detections from the GASKAP-\hi\ pilot survey of the Magellanic clouds, where the points plotted here are restricted to Magellanic velocities (i.e. excluding the Milky Way foreground). Each candidate is coloured by its detection significance, with darker colours reflecting higher signal-to-noise. The background map is \hi\ column density from the GASS survey \citep{kalberla_haud_2015_gass}.}
  \label{fig:gaskap_pilot_mag_detections}
\end{figure*}

A recent breakthrough, highlighting the promise of deeper integrations with SKA-mid, is the high number of \hi\ absorption sources detected by the GASKAP-\hi\ Pilot Survey in the Magellanic System, as shown in Figure~\ref{fig:gaskap_pilot_mag_detections} from \citet{dempsey2026}. Initial results are indicating that, with its 200 hours of observing time per position, the full GASKAP-\hi\ untargeted survey should reach an absorption source density of $\sim 11\,$source\,per\,deg$^{2}$, a 4--10 fold increase compared to previous targeted surveys \citep{jameson2019,dickey2000}.

\citet{chen2025} presented initial results for absorption spectra located in the outskirts of the SMC (10 targets) and LMC (18 targets) where there is less confusion on the line-of-sight. They find low CNM temperatures ($T_s \sim 27$\,K and $\sim 24\,$K in the SMC and LMC respectively). The SMC sample shows rather low CNM mass fraction ($f_{\text{CNM}} \sim 6\%$) while the CNM abundance seems to be slightly higher and more variable in the LMC ($f_{\text{CNM}} \sim 21\%$) with a line of sight even showing no WNM at all ($f_{\text{CNM}} \sim 100\%$). 
Absorptions across the Magellanic Bridge are also detected both in GASKAP-\hi\ and MALS where \cite{morelli2025} reported high significance (S/N $\sim 10$) measurements. Initial results suggest even lower spin temperatures across the Bridge ($T_s \sim 23$\,K) compared to what was found in previous studies \citep{jameson2019}.

The high spatial resolution and sensitivity of SKA pathfinders are also showing the potential of mapping the thermal properties ($\tau$, $T_s$) of extragalactic \hi\ using extended continuum background sources. For example, \citet{park2026} used this technique to probe the CNM across the extreme star-forming region 30 Doradus in the LMC on $\sim$7\,pc scales using ASKAP data. The SKA will only improve upon these numbers, potentially opening access to resolved CNM mapping on $\sim1\,$pc scales in the Magellanic system.

Moving towards other extra-galactic systems, using the LGLBS data \citet{pingel2024} and \citet{stelea2026} reveal two and three sight-lines respectively with localized \hi\ absorption ($T_s=32\pm6$\,K; $T_s=30$--50\,K) in the dwarf irregular galaxies NGC 6822 and IC10, demonstrating the feasibility for analyzing resolved thermal properties of the CNM outside of the Magellanic Clouds. Both of these systems have low metallicity similar to the SMC ($\sim$20--30\% Solar) and possess correspondingly low CNM mass fractions, consistent with the theoretical expectation for low metallicity environments that lack fine-structure metal line coolants \citep{bialy2019}. However, recent high resolution simulations from \citet{kim2024} show that the thermal balance in low metallicity environments also depends on second-order processes, such as reduced attenuation of the FUV by dust leading to an increased efficiency of photo-electric heating. From these initial LGLBS results one could hope that substantially deeper integrations allowed by SKA-mid could provide multiple absorption measurements within a single galaxy.

\subsection{Galactic halos}

\label{sec:halos}

\begin{figure*}
  \centering
  \includegraphics[width=0.9\linewidth]{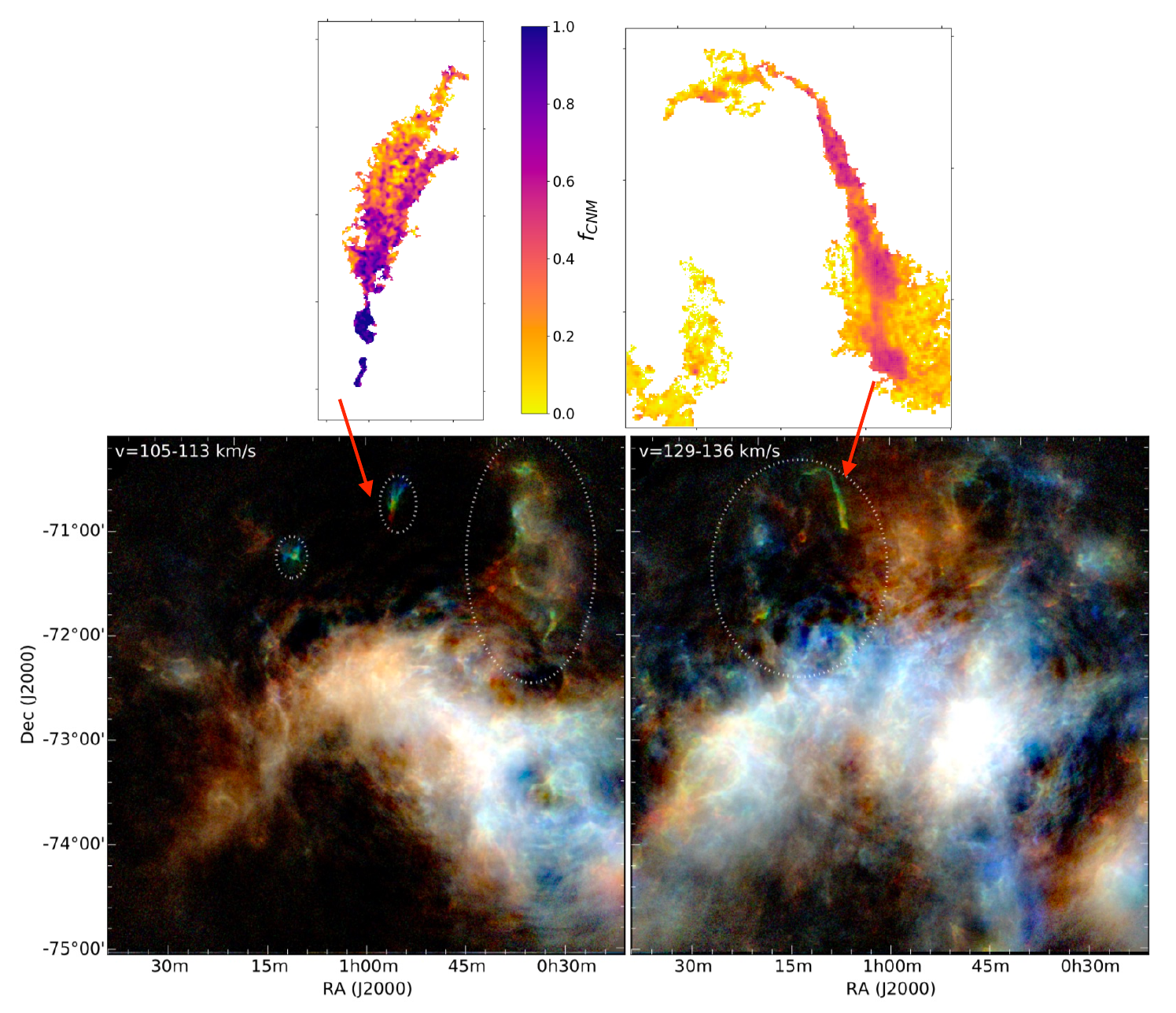}
  \caption{\hi\ clouds in the outskirts of the SMC. {\bf Top:} Two halo clouds for which we show the CNM fraction, $f_{\rm CNM}$, obtained from spectral segmentation of GASKAP-\hi\ 21\,cm data using ROHSA \citep{bucklandwillis2025}. {\bf Bottom:} GASKAP-\hi\ 21\,cm velocity channels of part of the SMC highlighting gas in the periphery of the galaxy \citep[see][for details]{mcClure-griffiths2018}.}
  \label{fig:SMC-halo}
\end{figure*}

\hi\ is a powerful tool for tracing gas transported by feedback and accretion processes in and out of galaxies. 
With \hi\ we see the neutral gas component of this cycle, often in the form of clouds or complexes with velocities incompatible with Galactic rotation, classified as intermediate or high velocity clouds \citep[I/HVCs;][]{wakker2004,martin2015,putman2012}. IVCs and HVCs around the Milky Way have been catalogued and mapped \citep{wakker_1991,deheij2002,putman2002,westmeier2018} using large \hi\ surveys \citep{hartmann1997,barnes2001,hi4pi}. IVCs are generally thought to trace the return flow of the galactic fountain \citep{shap1976,breg1980,houck1990}, a cycle describing interstellar matter being ejected, interacting with the corona, condensing and re-accreting onto the galaxy \citep{armillotta2016,li2023,barbani2023}.

Recent MeerKAT observations of such clouds located around the Galactic centre suggest the progressive evaporation of \hi\ as ejected clouds move away from the galactic centre in an outflow \citep{noon2023}. In such a scenario, simulations show that clouds can fragment and reform their cool material, and that magnetic fields support the survival of denser structures \citep{joung12,heitsch2009,heitsch2022,jung2023}. In contrast to IVCs, many HVCs have low metallicity, and are thought to represent gas accretion from the circum-galactic medium. The HI emission observations are beginning to be detailed enough to compare the cloud simulations to observations and identify the relevant physical processes \citep{porter25}. In the Milky Way HVCs have a sky covering factor of 15--35\% for column density limits of $2 \times 10^{18}$--$7 \times 10^{19}$ cm$^{-2}$ \citep{richter2017}.

Direct \hi\ absorption measurements of I/HVCs remain scarce, with only a few reported detections, e.g., in the Low-Latitude Intermediate-Velocity Arch 1   \citep[$T_s\sim75$\,K,][]{vujeva2023} and a detection of $T_s=47$\,K in the HVC complex H \citep{wakker_1991}. This is primarily due to the limited surface filling factor of bright radio sources and bright halo cores and substructures \citep{vanwoerden2004}, a situation that will improve dramatically with SKA-mid.
On the other hand, the fact that these clouds are relatively isolated in 21\,cm PPV space facilitates the modelling of their emission. With the advent of higher resolution and more sensitive observations, it has become possible to carry out phase decompositions of these clouds. IVCs and HVCs observed at 21\,cm are thus fantastic laboratories in which to study the dynamical evolution of matter in the disk-halo circulation, the thermal state in which it enters the disk, and the dynamical impact it has on the general ISM \citep{Miville-Deschenes:2017}. 
For example, using GBT and DRAO 21\,cm data, \citet{marchal2021} showed that a network of cold filaments has formed efficiently in Complex C --- the largest HVC on the sky, located $10\,$kpc in the Milky Way halo (see Fig.~\ref{fig:DHIGLS}).
Such halo clouds have also been detected in M31 \citep{thilker2004} and now, with GASKAP-\hi, in the Magellanic Clouds \citep{mcClure-griffiths2018}.
\citet{bucklandwillis2025} investigated the cold to warm \hi\ fractions of three such clouds in the SMC outskirts. They found two clouds had significant cold \hi\ fractions in regions closest to the galaxy (up to $100\%$ in one case), suggesting inflow/outflow processes acting upon the \hi, with the other cloud likely forming through chimney or shell-like expansion on the edge of the galaxy (see Fig.~\ref{fig:SMC-halo}).


\section{A new era with SKA: toward a unified view of the multi-phase ISM}

\label{sec:SKA-observing-program}

\subsection{The SKA leap}
\label{sec:why_ska}

The scientific drivers outlined above are currently actively addressed by the use of a powerful ecosystem of single dishes and interferometers. However, SKA will bring a monumental leap forward enabled by the simultaneous combination of several capabilities:

\begin{enumerate}
\item \textbf{A dramatic increase in absorption grid density with matched emission and absorption.} Surveys such as GASKAP-\hi\ and MALS have demonstrated the power of untargeted absorption surveys, already increasing the number of detections by orders of magnitude. The SKA will push this into a new regime by providing higher sensitivity, emission and absorption at matched angular resolution, and far larger samples, enabling thermodynamic inference (spin temperature, optical depth, and phase fractions) as a function of Galactic environment and external galaxy properties.

\item \textbf{Imaging fidelity, large dynamic range and high resolution for diffuse emission.} A major challenge in \hi\ science is the simultaneous imaging of diffuse, low surface-brightness emission and compact cold structures. The SKA will provide improved imaging fidelity and sensitivity, enabling reliable characterization of diffuse WNM/UNM, filamentary CNM networks, and their interfaces, down to physical scales (mpc) that are almost unexplored. This is crucial for turbulence studies, fiber imaging, and disk--halo studies.

\item \textbf{Unified continuum, polarization, and spectral-line surveys.} The SKA will enable highly complementary datasets: \hi\ emission and absorption, continuum mapping, RM grids, Faraday tomography, OH, Zeeman, RRL, and pulsar observations. This unification will allow magnetic field and ionized gas structure to be directly linked to \hi\ phase structure, providing a uniquely complete view of the multi-phase ISM.

\item \textbf{A Southern hemisphere legacy dataset.} Many of the most influential higher-resolution datasets of the Milky Way \hi\ (GALFA-\hi, CRAFTS, DHIGLS) are Northern-hemisphere limited. The SKA will deliver comparable or superior products over the Southern sky, enabling global studies of the Milky Way and its nearest external analogues.
\end{enumerate}

Thus, a SKA legacy survey of the diffuse \hi\ would be the enabling dataset for a new generation of multi-phase ISM studies in which the multi-scale thermal structure, turbulence, and magnetization can be inferred statistically and spatially, rather than along isolated lines of sight.

\subsection{Data analysis leap: unified exploitation of \hi\ emission and absorption}
\label{sec:abs_em_unified}

A major observational frontier in \hi\ astrophysics is to measure not only the distribution and kinematics of atomic gas, but also its thermodynamic structure through optical depth and spin temperature. Classical absorption/emission studies provide direct constraints on these quantities, but historically remained limited to sparse lines of sight due to the scarcity of suitable background continuum sources. This situation has changed rapidly with ASKAP and MeerKAT, which have demonstrated the feasibility of wide-area ``absorption grid'' surveys delivering multiple detections per square degree \citep[e.g.][]{nguyen2025,gupta2025}.

At the same time, fully sampled emission cubes contain rich spectral and morphological information that can be exploited to infer phase structure and identify cold filamentary networks.
The emission-based methods described in section~\ref{sec:phases-from-emission} illustrate and quantify a clear physical connection between \hi\ morphology and phase structure, and unlock new avenues for improving future phase separation techniques by making use of both spectral and spatial information to decompose \hi\ in PPV space.

Looking towards the \hi\ science with SKA-mid, there is a growing consensus within the community that the decomposition strategies highlighted here (absorption-emission, radiative transfer modeling, and phase separation of fully sampled emission data) provide complementary insights and should therefore be analyzed jointly to better understand the properties of the neutral ISM.
While SKA pathfinders are paving the way for constructing grids of absorption-emission pairs, there is tremendous potential to merge and unify decomposition strategies to fully exploit the unprecedented datasets that will be produced by SKA-mid.
Achieving this integration will require innovative research that builds on the decade of progress achieved on both fronts, combining spatially coherent phase decomposition with joint emission-absorption analysis in a Bayesian framework to infer realistic ranges for unobserved properties (e.g., thermal pressure), thereby delivering a truly multi-scale, multi-phase view of the neutral ISM in the Milky Way and nearby galaxies.

\subsection{A tiered SKA legacy survey concept: from kpc context to mpc physics}
\label{sec:tiered_concept}

The broad range of science goals outlined above cannot be fully addressed by a single survey design optimized for one spatial scale. Instead, we advocate a tiered strategy that naturally connects the large scale Milky Way and nearby galaxies context to targeted high-resolution studies of cloud formation in the solar neighborhood. Such a strategy also provides a natural roadmap for SKA development from early array configurations (AA*) to the full AA4 capability.

\paragraph{Tier~0: a hemispheric context survey of diffuse \hi.}

The first tier is a wide-area survey optimized for sensitivity to large-scale diffuse emission in order to reconstruct the global distribution of \hi\ across the sky. The main deliverable is a uniform, well-calibrated spectral-line cube covering a large fraction of the Southern sky, with sufficient angular resolution to separate major filamentary structures and Galactic features, and sufficient spectral resolution to trace cold gas kinematics.
The strength of such \hi\ survey lies in its uniformity, sensitivity, and ability to be combined with total-power data to recover large-scale emission.
The value of an all-sky \hi\ data cube resides in the understanding of interstellar structure formation and evolution in many different contexts and scales: 1) disk--halo interface, 2) Galactic plane including the Galactic center, and 3) Local Bubble, including the reconstruction of the interstellar volume within 1\,kpc from the Sun using Gaia data. 

For fully-sampled emission, the goal is to obtain a survey complementing GALFA-\hi\ and the upcoming CRAFTS survey done with FAST \citep{zhang2019} --- the best large-scale surveys in the Northern sky. SKA-mid AA4 can achieve the performances of GALFA-\hi\ \citep[surface-brightness sensitivity of $150$\,mK in a $\sim1.0$\,km\,s$^{-1}$ channel, and a spatial resolution of $\sim4'$, see][]{peek2018} for a very modest integration time of 2m20s per 1\,deg$^2$ pointing. GALFA has already observed down to DEC=$0^\circ$, and CRAFTS should extend this to DEC=$-15^\circ$ so an SKA-mid survey would need to cover only DEC$<-15^\circ$, corresponding to $\sim 75\%$ of the southern hemisphere (or $1.5\pi$). Such a survey could be done in less than 600 hours with AA4\footnote{For AA*, the integration time to match GALFA-\hi\ is about 10 minutes per pointing, or 2250 hours to cover $1.5\pi$.}, and maybe done commensally with other all-sky surveys conducted with SKA-mid. Any increase in the observing time would translate to increases in the angular or spectral resolution of the survey. Even with moderate observation time, such survey would provide a large number of absorption sources across the sky. As a comparison, SKA-mid AA4 can achieve $\sim10$ detections per deg$^2$, like in the deepest GASKAP-\hi\ full survey fields, with a 1h integration time\footnote{Excluding baselines $<250$\,m, SKA-mid AA will achieve 1$\sigma$ sensitivities of $\lesssim 1.3$\,mJy/beam in a 0.3\,km\,s$^{-1}$ channel, where the exact sensitivities are dependent on choices of weighting and tapering.}.

Combined, these three \hi\ surveys would improve on HI4PI (the current best all-sky survey) by close to an order of magnitude in combined angular and spectral resolutions\footnote{in practice the combination of these three surveys would cover 95.3\% of the sky as CRAFTS is limited to DEC<$+65^\circ$.}. 
Such an all-sky survey would provide the \hi\ counterpart to all-sky dust emission and polarization surveys (WISE, IRAS, AKARI, Planck), UV maps (GALEX), and large-area optical surveys (e.g. Euclid). Such a fantastic legacy product would be used for decades for interstellar science and to mitigate Galactic foreground contamination for extra-galactic and cosmological studies to much greater precision. It would also serve as a great exploration data set to identify targets to observe at higher resolution.

\paragraph{Tier~1: linking Galactic and extra-galactic \hi\ exploration:}
The second tier is designed to connect \hi\ condensation and phase structure directly to star formation and galactic dynamics in external galaxies. The onset of star formation occurs on scales of a few parsecs, comparable to the characteristic scale of cold ISM structures. Modern JWST, ALMA, and optical surveys routinely probe these scales in nearby galaxies, and there will certainly be a SKA-mid program that will explicitly aim to deliver a matched-resolution neutral gas view \citep[e.g.][]{deBlok2015,Blyth2015}.

Ideally this tier would target a sample of nearby systems spanning a range in metallicity and star formation environments. 
We argue that such a nearby galaxy program should perform observations at high spectral resolutions ($\delta v \sim 0.5\,$km\,s$^{-1}$) to measure the \hi\ thermal properties (WNM and CNM). 
The SKA survey speed, unprecedented angular and spectral resolution, and sensitivity will enable simultaneous wide-field \hi\ emission and absorption measurements on physical scales of $\sim1--10\,$pc, providing direct comparisons with dust and star formation tracers, and enabling measurements of the thermal state and turbulence of atomic gas in environments where feedback and galactic dynamics differ from the Milky Way.

Moreover, such targeted pointings towards nearby galaxies, observed with deeper integration than proposed in the first tier, could be used for commensal Galactic \hi\ science on random regions of the Solar neighborhood. Such a strategy is already producing interesting results with GASKAP-\hi\ (the Milky Way foreground towards the Magellanic clouds; \citealt{nguyen2025}). For example, deeper integrations of 20 hours with SKA-mid AA4\footnote{a relatively short integration time compared to the 55 hours spent on each galaxy with MeerKAT in the MHONGOOSE survey \citep{deblok2024}.} would allow imaging of the \hi\ of the Local Bubble walls at 1\,arcsec over a $1^\circ$ field-of-view, corresponding to physical scales ranging from 1\,mpc to a few pc.
In addition, such a setup would push sensitivities to $\lesssim 300\,\mu$Jy/beam, providing a dense absorption grid of $\sim$25--30 sources of $>5$\,mJy per deg$^{2}$ \citep{condon1998}, with corresponding optical depth sensitivities of $\sigma_{\tau}<0.06$. For even moderately bright sources ($\gtrsim$100\,mJy; $\sim1.5$ per deg$^{2}$), this sensitivity would be $\sigma_{\tau}\sim10^{-3}$ --- sufficient to begin detecting the UNM and WNM. 

By combining nearby galaxies and Milky Way foregrounds in the same field of view, these targeted observations would allow us to sample a very large range of \hi\ scales, from mpc to hundreds of kpcs, providing an innovative way to study the multi-scale evolution of interstellar matter in galaxies. 

\paragraph{Tier~2: targeted high-resolution mapping of \hi\ structures in the Milky Way.}
The third tier targets the condensation process at small scales. The goal here is to observe the Galactic \hi\ at high spectral ($\delta v \sim 0.2\,$km\,s$^{-1}$) and angular resolution ($\sim 1$\,arcsec), and with enough sensitivity to obtain a deep absorption grid. This is a fundamental aspect of this project, required to obtain a large number of precise measurements of the CNM thermal properties, which can then be anchored to emission-based phase decomposition. High spectral resolution is necessary to isolate the CNM. It is also critical for robust Zeeman measurements, where sensitivity to small Stokes-$V$ signatures benefits from narrow channelization.

Each $1\,$deg$^2$ pointing would provide three orders of magnitude in interstellar scales, in the mpc to tens of pc range, allowing us to study the multi-scale structure of the ISM in unprecedented detail. Targets would range from high Galactic latitude cirrus clouds, the outskirts of molecular clouds, Galactic halo clouds, supernovae remnants and HII regions, allowing us to evaluate the impact of feedback on the formation of the CNM.
In particular we would look for HISA/HINSA, compressed gas in shells, filament networks, and magnetically aligned cold structures to understand the fundamental physics of the \hi\ evolution and test theoretical models of turbulence-driven condensation and magnetically aligned structures. The selection of targets will be informed by results from the other tiers and in coordination with other relevant surveys.
In this chapter we have presented results revealing the link between the ionized, neutral and molecular phases of the ISM. SKA offers unique opportunities to observe several interstellar tracers of these phases (21\,cm -- including Zeeman, OH, Pulsars, polarized synchrotron, RRLs) in a coherent way. We suggest that this tier should be conducted by combining these observational tracers, using both SKA-mid and SKA-low, on specific interstellar targets.

In particular, target choices should also be coordinated with Pulsar monitoring where TSAS can be observed in order to connect the \hi\ sampled at AU scales to the general context seen in emission at high resolution.
Targeted \hi\ monitoring observations against bright pulsars would vastly increase the number of measurements of temporal fluctuations in \hi\ absorption. If we require pulsars with a phase-averaged flux density of $>5$ mJy at 1.4\,GHz and a rotational period of $P_0>0.1$s (slow pulsars are necessary to achieve sufficient resolution in both pulse phase and frequency), there are $\sim$100 viable sources in the SKA sky. For a 5h total on-source time, the achievable optical depth sensitivities range from $\sigma_{\tau}(\nu)\sim0.01$--0.04 for $5$\,mJy to $\sim$0.0006--0.002 for the brightest $100$\,mJy sources\footnote{For SKA AA4, in tied array mode including all baselines $<10$ km, SEFD = $2kT_{\rm sys}/A_e = 2.1$ Jy. Assuming a sky temperature of 5\,K, an H{\sc i} peak brightness temperature of 100\,K, a total integration time of 5h, a pulse duty cycle of 0.1, and a binned channel width of 1\,km\,s$^{-1}$, gives $\sigma_{S_\mathrm{on}}(\nu)=0.6$--2.2\,mJy and $\sigma_{S_\mathrm{off}}(\nu)=0.2$--0.7\,mJy (where the range arises from the frequency-dependent variation in the H{\sc i} brightness temperature contribution to $T_{\rm sys}$).}. This is a step-change in both sensitivity and sample size.

A major goal of these Tier 2 observations would be to build templates of the $P-n$ diagram described in Figure~\ref{fig:caribou_results} for a variety of environments and physical conditions, with the aim of identifying the main physical scenarios that lead to gas condensation, cloud formation, and later gravitational collapse. This can be achieved with the high sensitivity, high resolution and high dynamic range provided by SKA, combined with tracers of the molecular and ionized phases of the ISM observed in a coordinated way on specific targets. Such a program would allow us to make an important leap forward in our understanding of the multi-scale, magneto-ionic physics of the interstellar medium. 


\subsection{Roadmap from AA* to AA4 and early SKA observations}
\label{sec:roadmap}

A key advantage of the tiered survey strategy is that it naturally maps onto the staged development of the SKA. Early array configurations (AA*) will already enable wide-area \hi\ mapping and absorption grid science at unprecedented sensitivity. These early surveys can deliver Tier~0 context products and initiate Tier~1 studies in selected regions. As the array evolves toward AA2/AA3 and ultimately AA4, the improved sensitivity and baseline distribution will enable progressively higher angular resolution and deeper targeted studies, opening the full Tier~2 extragalactic program.
In this framework, a SKA \hi\ survey would not be a single static project, but an evolving program that delivers immediate early-science return while building toward the full scientific vision of subpc-scale \hi\ thermodynamics and magnetized turbulence studies in nearby galaxies.


\subsection{Low spacings and combination with total-power data}
\label{sec:low_spacings}

A critical technical requirement for SKA \hi\ surveys is the accurate recovery of emission across a wide range of angular scales. Many of the key science drivers---diffuse WNM structure, disk--halo interfaces, HVC envelopes, and large-scale filament networks---depend on faithful reconstruction of low surface-brightness emission. This requires robust strategies for combining interferometric data with total-power measurements to recover the zero-spacing information. As already shown with GASKAP-\hi, the combination of interferometric data with single-dish GASS data is able to recover the multi-scale \hi\ emission. We would adopt the same strategy for SKA-mid observations.

For interferometric RRL and Faraday rotation surveys, maintaining sensitivity to large-scale structures will also be necessary. The combination of single-dish and interferometric data has long been standard in \hi\ studies, but the procedure is less common in polarization. \citet{Landecker2010} produced combined polarization observations in single-channel images, and \citet{ordog2025} demonstrated its importance in four-channel maps. In the SKA era, this will be necessary across thousands of frequency channels.
Broadband single-dish diffuse polarization observations are increasingly available from the Global Magneto-Ionic Medium Survey \citep{Wolleben2019,Wolleben2021,Sun2025,ordog2025dragons} for SKA Band~1 and Band~2, and will provide an important basis for combining SKA interferometric polarization products with large-scale information.


\section{Conclusion}
\label{sec:conclusion}

Neutral atomic hydrogen is the primary reservoir of interstellar matter in disk galaxies and represents the key interface between large-scale galactic dynamics and the dense molecular structures that ultimately form stars. Through its 21\,cm line, H\,{\sc i} provides an observational handle on both the spatial organisation and kinematics of diffuse gas, while absorption measurements uniquely constrain its thermal and density structure. Over the last decade, major progress has shown that the atomic ISM is far from smooth: it is highly turbulent, strongly structured into cold filaments and clouds, and closely linked to magnetic fields across a wide range of scales.

This new picture has emerged from the convergence of improved instrumentation and new analysis techniques. Large-area emission surveys have mapped diffuse structure with increasing fidelity, while ASKAP and MeerKAT have demonstrated that absorption-grid observations can deliver thermodynamic information over wide fields rather than along isolated sightlines. In parallel, modern inference methods---including machine-learning phase separation and Bayesian forward modelling---are now capable of extracting physical information from complex H\,{\sc i} data products in a statistically robust way.

SKA-mid will take the next decisive step by enabling a unified exploitation of emission, absorption, continuum, and polarization datasets at matched angular resolution over large sky areas. This will make it possible to map the distribution of optical depth, spin temperature, CNM fraction, and turbulent properties in a coherent framework, and to relate these physical quantities directly to the observed filamentary morphology. In addition, SKA-mid will open new parameter space for magnetic field studies through large samples of Zeeman detections and Faraday rotation measurements, and will connect the neutral medium to ionized gas via radio recombination lines and broadband polarization tomography.

The scientific return will be amplified through synergy with facilities probing complementary phases of the ISM and star formation, from ALMA and other millimetre observatories to JWST and major optical/infrared surveys. Within this multi-wavelength context, SKA-mid will provide the missing neutral-gas backbone required to build a physically consistent description of cloud formation, feedback, and baryon cycling in nearby galaxies.

Finally, the staged development of the SKA naturally supports a tiered legacy-survey approach, delivering early high-impact datasets while progressively extending toward deeper and higher-resolution studies. By combining wide-area mapping with targeted investigations of key regions, SKA-mid will establish a definitive reference dataset for multi-phase ISM physics, enabling quantitative tests of models of thermal instability, turbulent dissipation, and magnetic regulation of the atomic-to-molecular transition, in the Milky Way and nearby galaxies.


\bibliographystyle{abbrvnat-maxbibnames4}
\bibliography{chapter}

\end{document}